\documentclass[preprint]{aastex}

\shorttitle{WFPC2 Observations of Andromeda III}
\shortauthors{Da Costa, Armandroff & Caldwell}

\begin{document}

\title{The Dwarf Spheroidal Companions to M31: WFPC2 Observations of 
Andromeda III\footnote{Based on observations with the NASA/ESA {\it 
Hubble Space Telescope}, obtained at the Space Telescope Science Institute,
which is operated by the Association of Universities for Research in 
Astronomy, Inc., (AURA), under NASA Contract NAS 5-26555.}
}

\author{G. S. Da Costa}
\affil{Research School of Astronomy \& Astrophysics, Institute of Advanced
Studies, The Australian National University, Cotter Road, Weston, ACT 2611, Australia}
\email{gdc@mso.anu.edu.au}

\author{T. E. Armandroff}
\affil{National Optical Astronomy Observatory, P.O. Box 26732, Tucson, 
Arizona 85726}
\email{armand@noao.edu}

\and

\author{Nelson Caldwell}
\affil{Smithsonian Astrophysical Observatory, 60 Garden Street,
Cambridge, MA 02138}
\email{caldwell@cfa.harvard.edu}

\slugcomment{Accepted for Publication in the Astronomical Journal, July 2002}

\begin{abstract}

The {\it Hubble Space Telescope} WFPC2 camera has been used to image 
Andromeda~III, a dwarf spheroidal (dSph) companion to M31.  The resulting
color-magnitude (c-m) diagrams reveal for the first time the morphology 
of the horizontal branch (HB) in this dwarf galaxy.  We find that like 
Andromeda~I and Andromeda~II, and like most of the Galactic dSph companions, 
the HB morphology of And~III is predominantly red, redder indeed than that of 
both And~I and And~II despite And~III having a lower mean metallicity.  The 
And~III HB morphology is also somewhat redder than that of the Galactic dSph 
Draco, which has a similar mean abundance to And~III\@.  We interpret this 
red HB morphology as indicating that the bulk of the And~III population is 
$\sim$3 Gyr younger than the age of the majority of Galactic globular clusters. 
Nevertheless, the And~III c-m diagram does reveal the presence of a few blue 
HB stars, and a number of RR Lyrae variables are also evident in the data.  
This indicates that And~III does contain an ``old'' population of age 
comparable to that of the Galactic globular clusters.  There is no evidence, 
however, for any young stars in And~III despite a claimed association 
between this dSph and an HI cloud.  As was the case for And~II, but not And~I,
no radial gradient was detected in the And~III horizontal branch morphology.
The mean $V$ magnitude of the horizontal branch is 25.06 $\pm$ 0.04
leading to $(m-M)_{0}$ = 24.38 $\pm$ 0.06 for this dwarf.  And~III is then
$\sim$75 kpc from the center of M31, comparable to the Galactocentric
distances of Sculptor and Draco.  Comparison with standard globular cluster
red giant branches indicates a mean abundance for And~III of $<$[Fe/H]$>$ = 
--1.88 $\pm$ 0.11, the lowest mean abundance of any of M31's companions.  
This value, however, is consistent with the absolute magnitude -- mean 
abundance relation followed by dSph galaxies.  The same comparison yields an 
intrinsic abundance dispersion for And~III of $\sigma_{int}$([Fe/H]) = 0.12, 
a low value compared to And I and And II and to the Galactic dSphs of 
comparable luminosity to And~III\@.  If confirmed by future spectroscopic 
studies, this low value would suggest that And~III retained relatively little 
of the enrichment products generated during its evolutionary history.  The 
list of candidate variables reveals one definite and one probable Anomalous 
Cepheid variable stars in And III\@.  Such variables are common in Galactic 
dSphs, so their discovery in And~III is not unexpected.

\end{abstract}


\section{Introduction}

The Galaxy's dwarf spheroidal (dSph) companions show a surprising
diversity in the epochs of their {\it major} star formation episode(s):
for some galaxies that epoch occurred $\sim$15 Gyr ago while for
others it was as recent as $\sim$2--3 Gyr ago.  In addition,
the majority of the Galaxy's nine dSph companions have had 
extended star formation histories. These histories are also diverse,
with multiple distinct episodes of star formation in some
systems and approximately continuous star formation over a significant
fraction of the Hubble time in others (see, for example, \citet{GD98},
\citet{EG99} and \citet{MM98}).

This variety of star formation histories was initially hinted at by the
range of horizontal branch (HB) morphologies seen in the Galactic
dSphs.  For the majority of Galactic globular clusters, there is
a well established relation between metal abundance and HB morphology.
This relation arises from the fact that the globular clusters of the Galaxy,
especially those in the inner halo, are essentially coeval (e.g.\
\citet{LDZ94}).  Consequently, if the mean abundance of a stellar system
is known, then it is possible to predict the HB morphology expected if
the dominant stellar population has an age comparable to the majority
of Galactic globular clusters.  When this test is applied to the dSph
companions of the Galaxy, however, most have redder HBs than
predicted.  In other words a ``second parameter'' is needed to explain
their HB morphologies.

The prime candidate for the ``second parameter'' is age, since younger,
higher mass stars are cooler (redder) in their core helium burning
phase of evolution.  For the Galactic globular clusters, the 
interpretation of the second
parameter as age remains controversial, though there are pairs of globular
clusters of similar
abundance but different HB morphologies where an age difference is well
established, such as NGC 288 and 362 (\citet{GN90}, \citet{BF01}; see also \citet{ST99}, \citet{RY01} and the references therein).  
But for dSphs,
there is little reason to question the assertion that metal-poor
systems with strong red HBs have mean ages younger than the Galactic
globular clusters.  Indeed, the assertion is given credence by direct 
determinations of the
mean ages of Galactic dSphs, via measurement of the main sequence
turnoff luminosity.  For example, main sequence photometry of Carina
\citep{TSH96,HK98} clearly shows the turnoff of an old population
($\sim$15 Gyr), which provides the progenitors for the RR Lyrae and blue
HB stars observed in this dSph.  Also visible is the turnoff of the dominant
intermediate-age ($\sim$4--6 Gyr) population, which produces the
numerous ``red clump'' stars -- intermediate-age stars in the core
helium burning phase of evolution.  Other examples where direct
measurement of the main sequence turnoff connects a mean age younger
than that of the Galactic globular clusters with a red HB morphology 
include Leo II
\citep{MR96}, Leo I \citep{CG99}, and Fornax \citep{SHS98}.  Ursa
Minor, which has the blue HB expected for its low metallicity and the
assumption of a dominant old population, also fits this picture; main
sequence turnoff observations confirm that this dSph is
indistinguishable in age from the Galactic globular clusters \citep{OA85,MB99}.

While these recent observations have revealed the diversity of star formation
histories among the Galactic dSphs, there is no accepted explanation
for its origin, only hints.  For example, there is a
tendency for Galactic dSphs with younger mean ages to lie at larger
distances from the Galaxy \citep{VB94}.  Clearly, we need to study
additional dSphs if we are to understand the origin of this diversity.
The nearest dSph galaxies beyond the Galaxy's companions are the dSph
companions to M31.  The availability of WFPC2 on the {\it Hubble Space 
Telescope}
allows accurate photometry at magnitudes fainter than the horizontal
branch in M31's dSph companions.  We have used {\it HST}/WFPC2
to obtain photometry for stars in two of M31's dSph companions,
And I and And II \citep{DA96,DA00}.  The resulting color-magnitude
diagrams for these dSphs reveal strong red HB morphologies.  These
morphologies are considerably redder than that expected from the mean
abundances of the dSphs, when coupled with the assumption of a dominant
stellar population similar in age to the Galactic globular clusters.
Instead, the HB
morphologies indicate that the bulk of the population in And I and II is
younger than the Galactic globular clusters.  In addition, the And I and II
color-magnitude diagrams have been used to derive accurate distances, from
the mean magnitude of the horizontal branch, and to determine each galaxy's
abundance distribution, from the colors of the red giants (\citet{DA96,DA00}; 
hereafter Paper~I and Paper~II, respectively).

In this paper, we use similar {\it HST}/WFPC2 data and analysis techniques 
to those of Papers~I and II to study Andromeda III\@.  
And III is one of the least luminous ($M_V$ $\approx$ --10.2, \citet{NC92}), 
and is apparently the most metal-poor ([Fe/H] $\approx$ --2.0, \citet{AD93}), 
of the M31 dSph (and dE) companions.  It lies between And I and And II in 
projected distance from the center of M31\@.  \citet{AD93}, hereafter ADCS93,
have presented a 
ground-based color-magnitude diagram for And~III, which shows the dSph's upper 
giant branch.  These authors suggest that the fraction of intermediate-age
(age between $\sim$3 and 10 Gyr) population in this dSph is 
approximately 10 $\pm$ 10 percent.  
And~III is of additional interest because of the claim of \citet{BR00}
that this dSph is associated with a H~{\small I} cloud.  In most, but not all situations (cf.\ Sculptor, \citet{CB98}),
association with a H~{\small I} cloud is reflected in the stellar population
of the dSph, via the presence of young stars (e.g.\ Phoenix, \citet{HSG00}; 
LGS~3, \citet{AG97, BM01}).  

The remainder of this paper is split into three sections.  In \S \ref{O_Rsect}
the observations, photometry techniques and calibration processes are described.
Section \ref{results} presents the main results of the paper.  These include,
inter alia, the HB morphology, the distance of And~III from the center of M31,
the mean abundance and abundance dispersion, and implications for the age of
the bulk of And~III's stellar population.  The final section discusses the
results in the wider context of dSph evolution.

\section{Observations and Reductions} \label{O_Rsect}

Andromeda III was imaged with the WFPC2 instrument aboard the {\it Hubble Space
Telescope} on 1999 February 22 and again, at the same orientation, on 1999
February 26 as per our GO program 7500.  Both sets of observations consisted
of four 1200~s integrations through the F555W (``Wide-$V$'') filter and 
eight 1300~s integrations through the F450W (``Wide-$B$'') filter.  And~III
is a highly flattened system ({\it b/a} $\approx$ 0.4, \citet{NC92}) 
and so an orientation range was specified at the proposal Phase-II
stage to ensure the major axis of the galaxy lay approximately across the WF2 
and WF4 chips.  Further, the WFALL-FIX center employed was offset somewhat 
from the actual center of the dSph in order to ensure that all bright 
foreground stars (cf.\ ADCS93, Fig.\ 2) fell 
well outside the WFPC2 field-of-view for the orientation range specified.  
The actual observations placed the center of And~III on the WF2 chip, 
approximately 18$\arcsec$ from the WFPC2 pyramid apex.  The orientation 
of the observations differed only by about 4 degrees from that required to 
place the And III major axis along the WF2--WF4 diagonal (U2, V2 axes).  This 
difference is within the uncertainty of the determination of the And~III 
major-axis position angle.  As for our previous observations of And~I and 
And~II, the second set of And III observations was offset from the first 
by a small 
amount, nominally 9.5 WF pixels in both $x$ and $y$.  This facilitates 
distinguishing real stars from instrumental defects, such as hot pixels, and 
allows an assessment of the errors in the photometry. 

The raw frames were processed via standard STScI pipeline procedures.  The 
processed frames were then separated into the images for each individual CCD,
multiplied by the appropriate geometric distortion image as supplied by STScI, 
and trimmed of the vignetted regions using the boundaries recommended in the
WFPC2 Handbook.  The locations of bright stars were then measured
on each set (one for each combination of filter, position and CCD) to
ascertain whether there were any systematic changes in position
during the sequence of observations.  None were found.  The individual
images for each set were then combined using the $gcombine$ task
within the STSDAS package.  As for our previous WFPC2 data sets, a value
of 0.10 was adopted for the $gcombine$ parameter $snoise$ as the best 
compromise between
cosmic-ray rejection and ensuring that the magnitudes of bright stars on the
combined frame are not significantly different ($\lesssim$0.02 mag)
from the average of the values from the individual frames.  Further, for these
data, it was apparent that the background sky level is significantly lower
on the second of the two exposures per orbit.  Thus, in using $gcombine$, 
additive offsets were applied to bring the set of frames to the same 
background level before initiating the rejection process.

A mosaic made from the combination of four 1200~s 
F555W frames is shown in Fig.\ \ref{wfpc2pic}.  As for the And~I and II data, 
the HST/WFPC2 combination completely resolves this dSph; indeed the And~III 
stars are relatively uncrowded.  There is also no indication 
of any star clusters on this image (nor are there any candidates from 
ground-based imaging), a result that is not surprising given the relatively
low luminosity of this dSph.  The strong flattening of And III is also
evident in this image -- it is the reason for the comparative absence of
stars, for example, in the parts of the WF3 chip furthest from the camera 
apex.

\placefigure{wfpc2pic}
\notetoeditor{Please reproduce this figure "upside-down" with respect to
the way the postscript file prints -- i.e.\ so that the arrow is approximately
up and the line is to the left.}

\subsection{Photometry}

As for And~I and And~II, the relatively uncrowded nature of the And~III 
frames means that we can employ the techniques of aperture photometry.  
We note first, however, that as in Paper~II we have not used any data
from the PC1 frames.  This is because the small area relative to the WF
CCDs, together with its location on And~III's minor axis, results
in only a small addition to the total number of stars from the PC data.  
Further, there is
also a comparative lack of ``bright'' stars on the PC frames with which to 
determine aperture corrections.  Consequently, all subsequent
discussion applies only to the three WF CCDs, though nevertheless, we
will include the PC in our subsequent full analysis of the And~III
variable star content (cf.\ \citet{BP01}).

The photometry procedures adopted were identical to those described
in Paper~II\@.  The initial image-center coordinate lists were generated
with the {\it daofind} routine within IRAF/DAOPHOT\@.  Aperture photometry 
was then carried out using a 2 pixel radius aperture with the ``sky'' taken 
as the mode of the pixel values in an annulus of inner radius 5 and outer 
radius 15 pixels.  Typically a dozen or so of the brightest, most isolated
stars distributed across each WF CCD were then used to determine the 
aperture corrections -- the difference between the 2 pixel measurement aperture 
and the 5 pixel standard aperture adopted by \citet{H95B}, hereafter H95.  
The values for the individual stars were then averaged
to form a single correction for each filter/position/WF CCD combination.  
These values are typically uncertain by 0.01 -- 0.02 mag.  The corrected 
photometry lists for the two positions and a given filter were then compared.  
There were no indications of any systematic differences
and so a single magnitude was generated by averaging the two measurements.  
Stars detected on only one frame were discarded.  Similarly,
the combined photometry lists for the F555W and F450W filters were 
matched to produce F450W--F555W colors.  Again stars not detected in both
filters were ignored.  At this stage corrections for exposure time, 
gain factors and zero-point were also applied to place the photometry on 
the H95 system.

As for Paper~II, to minimize the effects of crowding, any star whose center
lay within 5 pixels of the center of another star was removed from the
photometry lists.  Given the relatively uncrowded nature of these data, the
number of stars removed by this process was small (typically 5\%) but it did
reduce the scatter in the c-m diagrams.
Finally, the remaining ``stars'' were visually inspected on the 
four-exposure-combined F555W images and a few objects, 
mostly marginally resolved galaxies, removed.  The final And~III sample
then consists of 1630 stars from the three WF frames.

\subsubsection{Charge-Transfer Efficiency Corrections} \label{cte_sect}

The WFPC2 CCDs are known to suffer from poor charge-transfer efficiency
(CTE) and this effect is gradually worsening with time (e.g.\ \citet{WHC99} and
references therein).  Without correction, poor CTE results in systematically
fainter magnitudes for stars that undergo the largest number of charge
transfers (i.e.\ those with large $x$ and $y$ pixel coordinates). 
As was the case for Papers~I and II, we have therefore chosen to correct 
our photometry for CTE effects.  Specifically, noting that the analysis 
carried out by \citet{WHC99} contains data obtained as recently as
$\sim$2 weeks prior to our And~III observations, we have used equations 
1a, 1b and 3 of 
\citet{WHC99} to correct the raw photometry for CTE effects.  
We then regenerated the final averaged, crowding corrected, edited and 
calibrated photometry list by repeating the steps outlined in the previous 
section.  For our observations, the \citet{WHC99} correction process makes a 
typical And~III red horizontal branch star near the center of a WF frame
brighter by about 0.08 mag and bluer by about 0.02 mag relative to the
uncorrected photometry.  The corrections are about twice as large for 
horizontal branch stars with large $x$ and $y$ coordinates.  Application of 
the \citet{DO00} CTE correction equations gives similar results.

\subsubsection{Photometric Errors}

The CTE corrected data can now be used to quantify the photometric errors
as a function of magnitude.  These photometric errors are those that 
result from the measurement process and are distinct from any systematic 
uncertainty in the data.  The latter uncertainty arises, for example, from 
errors in the zero-point calibration and in the CTE correction process.
To calculate the photometric errors we compared the photometry for the
two separate pointings using only the stars in the final sample.  The 
results of this process are given in Table \ref{error_tab} which lists, for 
the specified magnitude bins, the average error for the mean of the two 
measures (magnitude and color).  As was found in Papers I and II, for these 
uncrowded stars the photometric errors, particularly at fainter magnitudes,
are essentially those expected on the basis of photon statistics.  At the
brightest levels, however, the errors asymptote to a limit of $\sim$0.015 mag,
again as was seen in the previous analyses.  This limit presumably represents
the fundamental limit on the photometric precision for these data, given the 
effects of flat-fielding, dark subtraction and the frame combination 
process adopted. This limit does not affect the interpretation of the data 
in any way.  The values in Table \ref{error_tab} are in fact about 20\% larger 
than the equivalent values for the And~II photometry of Paper~II, despite 
similar exposure times.  This difference has its origin in the fact that the 
background sky values for the And~III frames are a factor of approximately
two higher.

\placetable{error_tab}

In the process of comparing the photometry for the two pointings, a small
number of stars were identified as having magnitude or color differences
larger by 3.5$\sigma$ or more than the mean difference for their magnitude
(which was always close to zero).  Based on the experience of Papers I and II,
it is likely that the majority of these stars with discrepant photometry
are in fact variable stars, principally RR Lyraes.  These stars will be
discussed further in $\S$ \ref{rrvar}.  Figure \ref{cmd1} shows the 
CTE corrected color-magnitude
diagram for the final And~III sample on the photometric system of H95.
The 22 candidate variable stars are plotted with a different symbol.

\placefigure{cmd1}
\notetoeditor{Please print this figure across two columns}

\subsubsection{$V$ and $B-V$ Zero-Points}

In Paper~II we used a set of ground-based deep $B$ and $V$ images of And~II 
to investigate the zero-points of the transformations from F555W to $V$
and F450W to $B$ given by H95.  It was found that the ground-based $V$ 
photometry agreed well with the WFPC2 F555W photometry 
when transformed to $V$ as
prescribed in H95.  For the $B$ magnitudes, however, the ground-based data
indicated a zero-point shift of 0.055 mag relative to the H95 F450W to $B$
transformation, in the sense of fainter $B$ magnitudes.  This offset (and the
lack of an offset in $V$) was confirmed by analyzing F450W and F555W
observations, obtained from the HST archive, of the $\omega$ Cen standard field 
(Paper~II).  The lack of an offset in $V$ and the presence of one in $B$ is
not too surprising given that the H95 F555W to $V$ transformation is empirical
while that for F450W to $B$ is not (see Paper~II).

For And~III, however, we do not have high quality well-calibrated deep $B$ 
and $V$ images with which to perform a similar zero-point investigation.  
Nevertheless, as a check on our photometric zero-point in $V$, we have compared
our WFPC2 photometry with that obtained by ADCS93 who used the KPNO 4-m 
telescope prime-focus CCD camera.  To carry out this comparison, the WFPC2 and 
KPNO 4-m images were inspected and a number of stars which could be reliably 
measured on the ground-based images selected.  These stars, which are 
distributed over the WF frames, are all at or near the top of the
And~III red giant branch.  For the 19 stars in the comparison, the
mean value of $V_{WFPC2}-V_{ADCS93}$ is --0.011 mag with a standard
error in this mean of 0.013 mag.  Excluding the most discrepant star,
the mean value of $V_{WFPC2}-V_{ADCS93}$ becomes --0.004 $\pm$ 0.011
mag.  We then conclude that the WFPC2 And~III F555W magnitudes have been 
reliably transformed to $V$\@.  ADCS93 also observed And~III in $I$ rather than 
$B$ and so no direct check of the F450W to $B$ transformation is possible from
these data.  However,
the relation between the $(B-V)_{WFPC2}$ and $(V-I)_{ADCS93}$ colors for the
stars is consistent with that for metal-poor globular cluster red giants
{\it if} the additional 0.055 mag zero-point adjustment found in Paper~II
is also applied in transforming the And~III F450W magnitudes.  

We have also searched the HST archive for observations of the $\omega$ Cen 
standard star field near in time to the And~III observations.  Photometry, 
using exactly the same procedures, including CTE correction, as outlined above,
was obtained for 19 local photometric standards from \citet{AW94} on a 
F450W, F555W pair of frames taken on 1999 June 14.  The mean differences,
between the WFPC2 photometry and that of \cite{AW94}, and their standard errors 
are --0.016 $\pm$ 0.007 for $V_{WFPC2}-V_{Walker}$ and +0.009 $\pm$ 0.012 for 
$(B-V)_{WFPC2}-(B-V)_{Walker}$, again after applying the additional 0.055 mag
adjustment.  These results then support the approach adopted in Paper~II -- use
the H95 F555W to $V$ transformation without modification and adjust the
zero-point of the H95 F450W to $B$ transformation by 0.055 mag in the
sense of fainter $B$ magnitudes and hence redder colors.  Consequently, we
have also adopted this approach for transforming the And~III photometry to
$B$ and $V$\@.  The resulting $(V, B-V)$ c-m diagram, again indicating the 
candidate variables, is shown in Fig.\ \ref{cmd2}.

\placefigure{cmd2}
\notetoeditor{Please print this figure across two columns}

\section{Results}   \label{results}

\subsection{The Color-Magnitude Diagrams of And III}

In general, the c-m diagrams for And~III presented in Figs.\ \ref{cmd1} and 
\ref{cmd2} bear a strong similarity to those for And~I and And~II (Paper~II).
The general morphology is that of an old metal-poor population -- a
relatively steep red giant branch and a significant number of (core-helium
burning) horizontal branch (HB) stars which are predominantly red in
color.  With regard to the extent of contamination of these figures by 
non-And~III members, the situation is clearly intermediate between that for
And~I and for And~II (cf.\ Paper~II)\@.  Since And~III is also intermediate 
between And~I and And~II in its projected distance from the center of M31, the 
assertion that the principal contaminants of these c-m diagrams are (relatively
metal-rich) M31 halo red giants is again supported (cf.\ Paper~II)\@.  The 
degree of contamination of the principal sequences of And~III by non-members
is thus very limited and will not be considered further.
The most striking difference between the c-m diagrams of Figs.\ \ref{cmd1} and 
\ref{cmd2} and the equivalent diagrams for And~I and And~II (Paper~II) is 
the much 
less broad red giant branch for And~III\@.  In this respect and in terms of
the HB morphology, the And~III c-m diagram of Fig.\ \ref{cmd2} bears a strong
resemblence to that for the Galactic dSph companion Draco.
To illustrate this point, we show in Fig.\ \ref{draco} the Draco c-m diagram
from \citet{PS79} at the same scale as Fig.\ \ref{cmd2}.  Aside from the
larger sample in Fig.\ \ref{cmd2}, the degree of similarity between the
two c-m diagrams is self-evident.  We shall elaborate on this similarity in the
subsequent discussion.

\placefigure{draco}

\subsubsection{The Horizontal Branch Morphology} \label{hbmorph}

As is evident from Figs.\ \ref{cmd1} and \ref{cmd2}, the horizontal
branch morphology of And~III is predominantly red.  This is also the case
for And~I and And~II (cf.\ Papers I and II) as it is
for most of the Galaxy's dSph companions.  In the c-m diagrams
the red side of the And III HB is blended with the
red giant branch and there is no clear distinction between the
reddest of the core helium burning stars and the red giants at this luminosity.
We take this to be a consequence of the photometric errors in the 
F450W--F555W colors
($\sigma$(F450W--F555W) $\approx$ 0.065 at the HB level, cf.\ Table 
\ref{error_tab}) and of the lower sensitivity, relative to $B-V$, of
F450W--F555W to differences in effective temperature.

In papers I and II we used an HB morphology index {\it i} = {\it b/(b +r)}
where {\it b} and {\it r} are the number of blue and red HB stars,
respectively.  We can readily use a similar index for And~III although
the process of comparing the And~III index value with those for And~I and II
is complicated by the lower mean metallicity of And~III 
(cf.\ Sect.\ \ref{abund}).  This lower abundance means both the red giant
branch and the core-helium burning stars are notably bluer than for
And~I and II\@.  Thus the
red side of the color range used to quantify the number of red HB stars 
in And~I and II (F450W--F555W = 0.60) is clearly inappropriate for
And~III\@.  We adopt two approaches.  First, as was done for And~I and II,
we constructed a color histogram for the stars in the magnitude interval
corresponding to the horizontal branch, 24.85 $\leq$ F555W $\leq$ 25.25
in this case.  This histogram shows a decline redward
of F450W--F555W $\approx$ 0.50 and we adopt this as the limit beyond which
red giants substantially outnumber red HB stars.  We then take {\it r}
as the number of stars in this magnitude interval with
0.35 $\leq$ F450W--F555W $\leq$ 0.50, yielding {\it r} = 305.  For {\it b}
we take the same magnitude interval but use --0.05 $\leq$ F450W--F555W 
$\leq$ 0.25 plus the two bluer, fainter stars.  This gives {\it b} = 30
and thus {\it i$_{III}$} = 0.09 $\pm$ 0.02 where the error is calculated 
assuming
that {\it b} and {\it r} follow Poisson statistics.  The subscript ``{\it III}" 
is used to denote that this value comes from And~III specific color limits.
Excluding the candidate 
variables, the corresponding value of {\it i$_{III}$} is 0.08 $\pm$ 0.015. 
 
The second approach used is to adopt the same color limits as were used
to define the And~I and II red HB star numbers and correct for 
the red giant branch star contamination by interpolation in the number
of red giants above and below the horizontal branch.  This approach yields
{\it r} = 266 and thus with {\it b} = 30, {\it i} = 0.10 $\pm$ 0.02 in good
accord with the first estimate.
For And~I
and And~II we found values of 0.13 $\pm$ 0.01 and 0.18 $\pm$ 0.02,
respectively, indicating that And~III, despite being the lowest metallicity
system (cf.\ Sect.\ \ref{abund}), has the reddest HB morphology of the
three Andromeda dSph companions studied in detail so far.

We can also compare the And III HB morphology directly with that of 
Draco, via the $(V, B-V)$ data of Figs.\ \ref{cmd2} and \ref{draco}.  For
Draco we adopt {\it r}$_{BV}$ as the number of stars with 19.85 $\leq$ $V$
$\leq$ 20.25 and 0.40 $\leq$ $B-V$ $\leq$ 0.60 (cf.\ \citet{PS79}).  For
{\it b}$_{BV}$ we use the number of stars in the same magnitude interval, but
with 0.00 $\leq$ $B-V$ $\leq$ 0.30 plus the three fainter stars with 
$(B-V)$ $\approx$ 0.00.  Then, for the \citet{PS79} data, {\it r}$_{BV}$ = 67
and {\it b}$_{BV}$ = 9 yielding {\it i}$_{BV}$ = 0.12 $\pm$ 0.04 for Draco.
For And~III, we adopt the same color limits for {\it r}$_{BV}$ and a 
magnitude range (cf.\ Fig.\ \ref{cmd2}) of 24.8 $\leq$ $V$ $\leq$ 25.2.  This
gives {\it r}$_{BV}$ = 224 for the sample including the candidate variables.
For {\it b}$_{BV}$  we adopt the same magnitude interval, but use --0.3 $\leq$
$B-V$ $\leq$ 0.30 plus the stars with $B-V$ $\leq$ 0.25 and 25.2 $\leq$ $V$
$\leq$ 25.5, once more including the candidate variables.  Hence {\it b}$_{BV}$ 
= 20 and {\it i}$_{BV}$ = 0.08 $\pm$ 0.02 for And~III\@.  This value is not
appreciably altered if the candidate variables are excluded.  It is also more 
correctly interpreted as an upper limit on the true value of {\it i}$_{BV}$
for And~III, as there are, for example, a further 80 or so probable red HB
stars with $(B-V)$ colors between 0.65 and 0.70.

Comparison of this value of {\it i}$_{BV}$ with that for Draco indicates that
And~III's HB morphology is probably somewhat redder than that of Draco.
We will discuss the implications of this outcome below.  Here it is worth
noting that these results make the blue HB morphology of the Galactic dSph companion Ursa Minor look more and more unusual.  Horizontal branch 
morphologies are now available for
a number of M31 dSph companions (Paper I, II; this paper; Armandroff et al.\
2002, in prep.) and there are no examples of HB morphologies even remotely 
like that of Ursa Minor.  This is also the case for the Galactic dSph
companions, as has been known for some time. 

One further point deserves discussion here.  In Paper~I we showed that And~I 
possesses a radial gradient in its HB 
morphology; the blue HB stars are found relatively more frequently beyond 
that dSph's core radius.
Similar HB morphology radial gradients are also known to occur in at least two
of the Galactic dSph companions (Paper~I, \citet{HK99}, \citet{MS99}).
However, no such HB morphology gradient was 
found for And~II (Paper~II)\@.  It is therefore of interest to see whether
an HB morphology gradient is present in And~III\@.  \citet{NC92} give the
(geometric mean) core radius of And~III as approximately 76$\arcsec$ 
(120$\arcsec$ on the major axis, given that the ellipticity of And~III is
0.6, \citet{NC92}).  We have therefore divided our photometry sample, using
the appropriate elliptical boundary, into ``inside the core radius'' and 
``outside the core radius'' samples.  The former contains 1338 stars with
an average distance from the center of And~III of 0.6r$_{core}$, while the
latter sample contains 292 stars at an average radial distance of 1.3r$_{core}$.
The most distant star is 2.2r$_{core}$ from the center.  We also split the
sample, again along an elliptical boundary, into two groups that contain
approximately equal numbers of stars.  This split occurs for a major axis
radius of 82$\arcsec$.  In terms of r$_{core}$, the stars in the inner 50\%
sample have an average radial distance of 0.45r$_{core}$, while the outer 
50\% sample has an average radial distance of 1.0r$_{core}$.

For each of these samples we have calculated the And~III HB morphology index
{\it i$_{III}$} defined above.  The resulting values are 0.08 $\pm$ 0.02
and 0.13 $\pm$ 0.05 for the inside and outside the core radius samples,
respectively, and 0.07 $\pm$ 0.02 and 0.10 $\pm$ 0.02 for the inner and
outer 50\% samples.  While these values hint at bluer HB morphology at
larger distances from the center of And~III, in fact the differences are
not statistically significant: calculation of the {\it T$_{2}$} statistic
(cf.\ Paper~I) indicates a $\sim$15\% probability that the differences arise
by chance alone.  Thus And~III joins And~II in apparently lacking any
HB morphology gradient, in contrast to And~I\@.

\subsubsection{Variable Stars} \label{rrvar}

In Fig.\ \ref{cmd1} and \ref{cmd2} the majority of the stars that show
discrepant colors or magnitudes between the two sets of observations
lie on or near the horizontal branch.  They are therefore most likely to
be RR~Lyrae variables, as was found to be the case for And~I and II (Paper~I,
Paper~II).  To confirm this is also the case for And~III, we have 
investigated a sample of these RR~Lyrae candidate variables.  This involved 
carrying out aperture photometry, in an identical fashion to that described 
above, including CTE corrections, for the candidate stars on the individual, 
as distinct from the combined, WFC frames.  This leads, modulo measurements 
affected by cosmic rays, to two sets of 8 F450W and 4 F555W individual magnitudes, separated in time by $\sim$4.8 days.  The total duration of 
each set is $\sim$0.4d, with the F450W observations occupying about 0.25d.
These durations naturally limit the precision with which RR~Lyrae like
periods (i.e.\ typically ~0.55d for ab-type RRs) can be determined.

The first step in the analysis was to plot the individual magnitudes 
against the epoch of mid-exposure to ascertain whether the star's brightness
varied in a way consistent with timescales and amplitudes characteristic
of RR~Lyraes.  The vast majority of the sample of stars investigated here did
show such characteristics, and thus approximate periods were then estimated.
In estimating these periods often a choice was required for the number of
cycles between the two sets of observations.  This was usually made 
considering the probable amplitude of variation.  Further, frequently the F555W
observations were also included in the period determination process, under the
assumption that color variations are much smaller than those in brightness. 
Typical light curves from the sample of stars analyzed are shown in Fig.\
\ref{rr_fig}.  These light curves, and others not shown, identify these
stars as Type-ab RR Lyraes.  It is important to remember, however, that
we have not yet analyzed the data for all the candidate variables, nor have
we yet conducted a more thorough search for all possible variables.  As noted
in Papers~I \& II, significant differences in color or magnitude in the
data from the two sets of combined frames is a sufficient, but by no means
necessary condition for detecting variable stars.  We postpone to a 
subsequent paper a full investigation of the frequency of occurrence and
properties of RR~Lyrae stars on the horizontal branch of And III.
 
\placefigure{rr_fig}

In Fig.\ \ref{cmd1} and \ref{cmd2} there are two candidate variables 
which are clearly brighter than the HB, yet which have colors that are
not dissimilar to those of HB stars.  Such stars, if they are indeed
variables, are likely to be either Anomalous Cepheids (ACs) or Population~II
Cepheids.  

Anomalous Cepheids, so-called because they do not obey the
Period-Luminosity relations for either classical or Pop~II cepheids
(cf.\ Fig.\ 1, \citet{ZS76}) are common in dSph galaxies -- they are found 
in all such systems where variable star surveys have been carried out.
On the other hand, they are extremely rare in the globular cluster
population, with only one example known -- V19 in NGC~5466 \citep{ZK82}.
The pulsation characteristics of ACs require masses in the range 1--2
M$_\sun$ and relatively low metallicities (e.g.\ \citet{NZ75}).  The first of
these requirements then suggests that either ACs
are stars of intermediate-age (age $\lesssim$ 5 Gyr, \citet{DH75}),
or that they are older stars whose increased mass has been generated by mass
transfer in a binary system (e.g.\ \citet{RMS77}).

In contrast to ACs, Pop~II cepheids are relatively common in the Galactic
globular cluster population and are certainly rarer than ACs in dSphs
(e.g.\ \citet{NNL94}).  Pop~II cepheids are understood as low mass (old)
stars with low envelope masses that enter the instability strip either
when evolving from a blue HB towards the AGB, or on blueward excursions from
the AGB, or in the final blueward evolution at the end of the AGB phase.

We consider first the brighter variable candidate WF2-1398.  Photometry of
the individual data frames as described above for the candidate RR~Lyrae
stars confirms that this star is indeed a variable.  The most likely period
is 0.830d and the light curve for this period is shown in the upper panel
of Fig.\ \ref{ac_lght}.  For this light curve the (intensity weighted)
mean F450W magnitude is approximately 23.5 and the amplitude is of order 
1 mag.  
Using the distance modulus and reddening from Sect.\ \ref{dist_sect}, and 
assuming that for a star of this color F450W $\approx$ $B$, we derive 
$<M_{B}>$ $\approx$ --1.1 for this star.  With this mean luminosity and period
the star then lies squarely on the P-L relation for first overtone AC pulsators
in Fig.\ 5 of \citet{BP01}.  Similarly, the amplitude and period are
consistent with those for the ACs presented in Fig.\ 7 of \citet{BP01}.  
We conclude
therefore that this star is indeed an Anomalous Cepheid member of And~III,
joining those in And~VI \citep{BP01} as the first such stars discovered
beyond the Galaxy and its companions.  Given that ACs are common in the
Galactic dSphs, the discovery of at least one AC in And~III is perhaps not surprising,
though a similar analysis to that carried out here did not
reveal any promising AC candidates in either And~I or And~II\@. 
This may merely be a consequence of the lack, at the present time, of a 
detailed 
search for such variables in those dSphs (cf.\ Papers I, II).

The occurrence of ACs in And III, and their apparent absence in And I and II, 
confirms the trend from Galactic dSphs that such stars occur relatively more
frequently in lower luminosity dSphs (e.g.\ \citet{MM95}).
Lower luminosity dSphs also have lower 
mean abundances and so this trend reinforces the result that, regardless of 
the mechanism(s) that generates anomalous cepheids (e.g. mass transfer binaries 
or single younger stars), a low metallicity stellar population is a necessary 
prerequisite for the existence of such stars.

\placefigure{ac_lght}

The interpretation of the second candidate variable brighter than the HB,
WF2-1710, is, unfortunately, not 
as straightforward though the individual frame data do indicate that 
the star is genuinely variable.  Plausible light curves can be 
constructed for periods 
longer than $\sim$1 day, but shorter periods do not look viable.  The
light curve that results from an assumed period of 1.086d is shown in the
lower panel of Fig.\ \ref{ac_lght}.  From this light curve the mean F450W
($\approx$ $B$) magnitude is approximately 24.55 corresponding to  
$<M_{B}>$ $\approx$ --0.1.  At this period and absolute
magnitude, the star falls somewhat below the P-L relation for 
fundamental pulsator ACs in Fig.\ 5 of \citet{BP01}.  It is unclear how significant this discrepancy is given the lack of complete coverage of
the light curve.  The amplitude of variation is approximately 0.7 mag.
This value and the assumed period are not obviously discrepant in
Fig.\ 7 of \citet{BP01}.

On the other hand, if the period is as long as 2.17d, which is equally
consistent with the observations, then this star is much more likely to be
a Pop~II Cepheid (BL Her star) than an AC\@.  If this is the case then it
would be somewhat unexpected for, at least in the Galactic globular clusters,
Pop~II cepheids are found only in those clusters with strong blue HB
morphologies.  \citet{NNL94} in their Table 4 list 20 BL Her stars in
10 Galactic globular clusters and the Harris catalog \citep{WH96} gives HB 
morphology indices (B--R)/(B+V+R) for 8 of these clusters.  The average value 
of this morphology index is 0.88, on a scale where 1.00 is an entirely blue HB
morphology and --1.00 is an entirely red HB\@.  The other two clusters, 
$\omega$ Cen and NGC~6273, also have strong blue HBs: \citet{WH96}, for 
example, lists Dickens HB classifications for these clusters
of 2 and 1, respectively, on a scale where 0 is a pure blue HB and 7 is pure
red HB\@.  Thus in Galactic globular clusters at least, it does appear that a
strong blue HB morphology is required to produce BL Her type Pop~II cepheids.
Consequently, given the strong red HB morphology of
And III, the occurrence of a Pop~II cepheid in this system would be 
unexpected.  We prefer
then to interpret WF2-1710 as a second AC in And~III, though further observations are
clearly required to constrain more stringently the nature of this variable.

\subsubsection{The Giant Branch Intrinsic Color Width} \label{clr_wid}

One of the characteristics that separates the Galactic dwarf spheroidal 
galaxies from the Galactic halo globular clusters (with the exception of 
$\omega$~Cen and perhaps also M22 and M54) is the presence in the dSphs 
of an intrinsic range in the
abundance of the member stars.  This abundance range can be revealed, given
sufficiently accurate photometry of large samples of stars, by an
observed red giant branch color width that is in excess of the width 
expected from 
the photometric errors alone.  In Papers I and II we confirmed earlier 
suggestions, based on less precise ground-based photometry, of intrinsic
giant branch color widths in the M31 dSphs And I and And~II\@.  Indeed, as
discussed in Paper~II, And~I and And~II have significantly different
giant branch intrinsic color widths, with that for And~II being substantially
larger, despite the similar mean metallicities and luminosities of these two
dSphs.  

ADCS93 obtained ground-based $V,I$ photometry of And~III and used it, 
inter alia, to infer the presence of an intrinsic giant branch color width
in this dSph.
However, because they did not completely characterize their $V-I$ color 
errors, they were unable to fully quantify this intrinsic color width.  
By following the procedures outlined in Papers~I and II, we can use our
And~III WFPC2 data to investigate this question further.  The first step is 
to define a mean giant branch locus.
This was done by fitting a low order polynomial to the stars in the 
interval 21.9 $\leq$ F555W $\leq$ 23.9, excluding obvious outliers.  Then,
for each of the 60 stars in the interval 22.3 $\leq$ F555W $\leq$ 23.3, again
excluding outliers, we 
computed the difference between the F450W--F555W color of the star and the 
color of the mean giant branch at the F555W magnitude of the star.  The mean
of these differences is --0.001 mag (as expected) with a standard deviation
$\sigma_{obs}$(F450W--F555W) = 0.048 mag.  This value is not notably altered 
for modest changes in the magnitude range considered or by the inclusion or 
exclusion of the few outliers that lie near the giant branch.  It is, however,
considerably smaller than the equivalent values for And~I (0.068 mag using the 
corrected And~I photometry of Paper~II) and And~II (0.12 mag, Paper~II) for
the same part of the red giant branch (between approximately 1.7 and 2.7
magnitudes above the horizontal branch).  The difference between these 
residual color distributions on the red giant branch is illustrated in Fig.\  \ref{delta_col}, where we have plotted the distributions for And~I, 
II and III, normalized by the sample sizes of 114, 85 and 60 stars, respectively.  In each case the photometric errors are comparable 
(cf.\ Papers I \& II) so that the obvious apparent differences are a
reflection of significant intrinsic differences.

\placefigure{delta_col}

Using the data of Table \ref{error_tab}, the mean error in the And~III
F450W--F555W colors for 22.3 $\leq$ F555W $\leq$ 23.3 does not 
exceed 0.03 mag.  Subtracting this value in quadrature then yields 
$\sigma_{int}$(F450W--F555W) = 0.038 mag for the And~III giant branch.
This confirms the results of ADCS93 in that the And~III giant branch 
clearly does possess a small intrinsic color width.  We defer to $\S$
\ref{abund_spd} a discussion of the abundance spread implied by this intrinsic 
color width.
 
\subsubsection{A Population of ``Faint Blue Stars''?}  \label{fbs_sect}

In Papers~I and II we noted that the And~I and II c-m diagrams show
a population of ``faint blue stars'' -- stars with relatively blue 
colors ($B-V$ $\lesssim$ 0.5) 
that lie at least 0.5 mag fainter than the horizontal branch.
It was not possible to dismiss these objects as artifacts of large
color errors in subgiant photometry, and so they were considered a bone fide
population in both dSphs.  For And~I, since there is no compelling
evidence for any intermediate-age population (\citet{MK90}, ADCS93, 
\citet{TA94}), these faint blue stars were interpreted in Paper~I 
as blue stragglers, analogous to
those seen in many Galactic globular clusters.  On the other
hand, for And~II there is evidence for an intermediate-age population
(see the discussion in Sect.\ 3.5.2 of Paper~II)\@.  Thus the faint blue
stars in this dSph could consist of main sequence and main sequence turnoff
stars as young as perhaps 1.5 Gyr, in addition to a blue straggler
component (Paper~II)\@.  Interestingly, these faint blue stars occur at a
significantly higher frequency in And~II than they do in And~I (Paper~II)\@.

As for And~III, there is some evidence from the c-m diagrams of Figs.\
\ref{cmd1} and \ref{cmd2} that a similar population
of ``faint blue stars'' exists, although qualitatively the relative number of 
such stars appears smaller than for And~II, for example.  Further, since
the lower mean metallicity of And~III results in a bluer subgiant branch, it
is possible that at least some of these And~III ``faint blue stars'' are 
actually
the result of large errors in the photometric colors of subgiant stars.  
In particular, 
based on the data of Table \ref{error_tab}, at $V$ $\approx$ 26, the 1$\sigma$
repeatability error in $B-V$ is approximately 0.18 mag, so that subgiants
with $\gtrsim$2$\sigma$ blueward errors will have $(B-V)$ $\lesssim$ 0.5 mag.
Nevertheless, it seems unlikely that photometry errors could generate the stars
with $V$ $\approx$ 26 and $(B-V)$ $\lesssim$ 0.15 visible in Fig.~\ref{cmd2}.

We can quantify the relative number of faint blue stars in And~III by 
following the same procedures as in Paper~II, in which the number of faint
blue stars, hereafter $FBS$, is compared to the number of subgiants, 
hereafter $SG$, with similar $B$ magnitudes (since the completeness is set by
the F450W data at these magnitudes).  Noting that the HB in And~III
is approximately 0.1 mag fainter than that for And~II, and that the reddenings 
are very similar, we take $FBS$ for And~III as the
number of stars with 25.65 $\leq$ $V$ $\leq$ 26.1 and $(B-V)$ $\leq$ 0.40 mag,
where the adopted color limit was chosen to minimize potential large color
error subgiant contamination.
There are 16 such stars with a mean $B$ magnitude of 26.1.  
For $SG$ we adopt the number
of stars with 25.15 $\leq$ $V$ $\leq$ 25.45 and 0.60 $\leq$ $B-V$ $\leq$ 0.95;
there are 89 such stars and their mean $B$ magnitude is again 26.1, so that the
completeness should be similar to that of the faint blue stars.  The relative
proportion of faint blue stars, $FBS/(FBS+SG)$, is then 0.15 $\pm$ 0.04
for And~III; the error is calculated assuming Poisson statistics.  The
equivalent values for And~I and And~II (Paper~II) are 0.21 $\pm$ 0.03 and 0.31 $\pm$ 0.04, respectively.  Comparison with these values confirms the
subjective impression that And~III is relatively lacking in these faint blue
stars, especially with respect to And~II\@.  Indeed, using the $T_{2}$ 
statistic (cf.\ Paper~I) , there is a less than 1\% chance that the And~II 
and And~III
faint blue star proportions are drawn from the same underlying population.

Given that there is little evidence for an intermediate-age population in 
And~III (ADCS93), it seems appropriate to conclude that these faint blue 
stars are predominantly blue stragglers analogous to those in the Galactic 
globular clusters.

\subsection{The Distance of And III}  \label{dist_sect}

The mean $V$ magnitude of the 34 blue HB stars in Fig.\ \ref{cmd2}
with 24.9 $\leq$ $V$ $\leq$
25.2 and 0.05 $\leq$ $B-V$ $\leq$ 0.45 is 25.06 $\pm$ 0.015 (standard
error of the mean).  This value is unaltered for minor changes in the 
adopted color interval or if the candidate variables are excluded.  We then
adopt $V$(HB) = 25.06 $\pm$ 0.04 for And~III, where the error now
includes uncertainty in the aperture corrections ($\pm$0.02) and in the 
zero-point/CTE corrections ($\pm$0.03).  This apparent horizontal branch
magnitude is intermediate between the values given in Paper~II for And~I 
(25.23 $\pm$ 0.04) and And~II (24.93 $\pm$ 0.03).  

To convert this value of $V$(HB) into a distance to And~III, however, we need 
to adopt a reddening and a mean metal abundance.  For the reddening, we follow 
Paper~II and use the value of $E(B-V)$ given by the dust maps of \citet{SFD98} 
at the location of And~III\@.  This value is $E(B-V)$ = 0.055 $\pm$ 0.01 so 
that $A_{V}$ = 0.18 $\pm$ 0.03 mag.  The mean metallicity of And~III is the
subject of the next section; here we simply adopt the result $<$[Fe/H]$>$ 
= --1.88 $\pm$ 0.11.  In Papers I and II we adopted a distance scale  
specified by the relation
$M_{V}$(HB) = 0.17[Fe/H] + 0.82.  For consistency, we adopt this relation
once more\footnote{The relation is based on the horizontal branch 
models of \citet{LDZ90}.  More recent models \citep{DZLY00} have shown that 
the slope of the relation between $M_{V}$(RR) and [Fe/H] is
dependant on abundance and that, at fixed [Fe/H], the value of $M_{V}$(RR)
depends also on the horizontal branch morphology.  Nevertheless, the new models
suggest our adopted relation is not significantly in error, given the mean
abundance and HB morphology of And~III\@.}.  The absolute magnitude of 
the And~III horizontal branch is then $M_{V}$(HB) = 0.50 $\pm$ 0.02 leading to 
$(m-M)_{V}$ = 24.56 $\pm$ 0.05 and $(m-M)_{0}$ = 24.38 $\pm$ 0.06,
or a distance of 750 $\pm$ 20 kpc on our adopted scale.
This result is in extremely good agreement with the And~III distance modulus,
$(m-M)_{0}$ = 24.4 $\pm$ 0.2, derived by ADCS93 from the $I$ magnitude of the
red giant branch tip.  We note that the same (horizontal branch
absolute magnitude, abundance) relation used here also underlies the 
ADCS93 modulus determination.  

In Papers I and II the distances of the And dSph galaxies were compared 
to M31 using two somewhat different M31 distance moduli (on our
adopted scale):
24.40 $\pm$ 0.13 based on RR Lyrae and field red giant data, and 
24.64 $\pm$ 0.05 based on mean horizontal branch magnitudes for 8 M31
globular clusters (cf. Paper~II)\@.  Recently, however, \citet{DHP01} have
used the $I$ magnitude of the red giant branch tip from a large
sample of M31 halo red giants to derive an M31 distance modulus of 
$(m-M)_{0}$ = 24.47 $\pm$ 0.12 (783 $\pm$ 43 kpc).  The calibration employed
by \citet{DHP01} is similar to that adopted here.  This value is clearly close 
to one of our earlier adopted values.  It is also consistent with the 
results of \citet{HO98}, who found $(m-M)_{0}$ = 24.47 $\pm$ 0.07 from fits of 
theoretical giant branches to WFPC2 data for a number of M31 globular clusters.  The \citet{HO98} results are also on a scale similar to that
used here.  We will therefore adopt 780 $\pm$ 45 kpc as the distance to M31 
for comparison with our And~III results.   

Adoption of this M31 distance then places And~III 30 $\pm$ 50 kpc in
front of M31 along the line-of-sight.  The projected distance of And~III
from the center of M31 is $\sim$67 kpc and thus the true distance of And~III 
from M31 is $R_{M31}$ $\approx$ 75 kpc.  This value is intermediate between
those for And~I ($R_{M31}$ $\approx$ 45 kpc) and And~II ($R_{M31}$ $\approx$ 
165 kpc) and is comparable
to the galactocentric distances of the nearer (excepting Sagittarius) of
the Galaxy's dSph companions.  For example, Draco and Sculptor have 
$R_{G}$ $\approx$ 75 and 79 kpc, respectively.

\subsection{The Abundance of And III} \label{abund}

To determine the mean abundance of And~III we again follow the procedures
outlined in Paper~II, and fit standard globular cluster giant branches to
the data of Fig.\ \ref{cmd2}.  Since the distance scale we employ is
sensitive to
abundance, this process is necessarily iterative, but convergence was
rapid since an initial estimate of the mean And~III abundance is available
from ADCS93.  The process involves calculating mean $B-V$ colors, excluding
obvious outliers, for stars in a series of 0.2 mag wide $V$ magnitude bins.
There were six bins on the upper giant branch between $V$ = 22.3 and $V$ =
23.5 and two bins on the lower giant branch between $V$ = 25.3 and $V$ =
25.7 (i.e.\ fainter than the horizontal branch).  Briefly, the lower giant
branch bins offer the advantage of freedom from stars evolving from the 
horizontal branch and larger samples, but suffer from lower sensitivity
to abundance, larger photometric errors and possibly more contamination
from non-members.  On the other hand, the upper giant branch offers
better abundance sensitivity and smaller photometric errors even if the
sample sizes are reduced.  The possible influence of post-HB stars is also
a potential concern at these luminosities but has not proved to be a
problem (cf.\ Papers I and II).  

The abundance corresponding to the mean color in each bin was determined
by applying a calibration for that bin based on the standard globular cluster
giant branches.  These calibrations used the giant branch colors at the 
midpoint (in $V$) of each bin and the known abundances of the clusters 
taken from \citet{A89}.  These abundances are on the \citet{ZW84} scale.
Linear least squares fits were used for all except the three highest
luminosity bins where quadratic fits were employed.  The results of this
process are illustrated in Fig.\ \ref{gbcmd} which shows the mean ($B-V$)
colors and the standard globular cluster giant branches superposed on the
And~III data using $(m-M)_{V}$ = 24.56 and $E(B-V)$ = 0.055.  The six higher 
luminosity bins give very consistent results (despite the small sample size in 
some cases) and yield $<$[Fe/H]$>$ = --1.88 $\pm$ 0.04, where the uncertainty 
given reflects only the statistical uncertainty in the mean colors.  The two
lower luminosity bins are also consistent with each other and yield 
$<$[Fe/H]$>$ = --1.72 $\pm$ 0.07, where again the uncertainty comes from
just the statistical uncertainty in the mean colors.  This latter abundance 
is 0.16 dex higher than the value derived from the upper giant
branch.  The difference between abundances derived from the upper and
lower giant branches is reminiscent of that seen in the equivalent analyses
in Papers I and II\@.  For example, in Paper~II the lower giant branch
mean colors gave an abundance for And~II that was 0.17 dex higher than the
value determined from the upper giant branch.  As was the case in Papers I and
II, we prefer to adopt the mean And~III abundance derived from the upper
giant branch colors.  This is primarily because the globular cluster
standard giant branches are not as well determined at the lower luminosities,
which combined with the strong abundance sensitivity to small changes in
color at these luminosities, apparently results in this systematic offset.

\placefigure{gbcmd}
\notetoeditor{Please print this figure across two columns}

The error in the mean abundance arising from the statistical uncertainty 
in the mean giant branch colors is, of course, not the only source of
uncertainty.  We have also to consider the effects
of uncertainty in the $(B-V)$ zero-point, uncertainty in the distance
modulus, and in the abundance calibration process itself.  We take these
to be $\pm$0.03 mag, $\pm$0.05 mag and $\pm$0.05 dex, respectively, where 
the last
value is a typical rms deviation about the least squares fits.  The first
two are equivalent to uncertainties in abundance of approximately 
0.08 and 0.04 dex, respectively.  Thus, while the uncertainty in the $(B-V)$ 
zero-point is the largest single contributor, it is by no means dominant. 
The combined uncertainty is 0.11 dex.  The mean abundance of And~III from our 
WFPC2 observations is then $<$[Fe/H]$>$ = --1.88 $\pm$ 0.11 dex.
This value is in excellent agreement with the earlier value, 
$<$[Fe/H]$>$ = --2.0 $\pm$ 0.15, determined by ADCS93 from the $(V-I)$ colors
of the And~III red giant branch in ground-based data.  And~III thus retains 
its distinction as
the dwarf with the lowest mean abundance among the known dSph companions to M31.
With our mean abundance and the And~III luminosity, $M_{V}$ = --10.35 $\pm$ 
0.1, that follows from the apparent $V$ magnitude measured by \citet{NC92}
and the apparent modulus derived in the previous section, And~III remains
part of the well established correlation between mean abundance and 
luminosity for dSph galaxies (e.g.\ \citet{NC99}).

One final point deserves mention.  We have investigated whether there is
any evidence for the presence of a radial abundance gradient in And~III by 
computing mean giant branch colors for the radially selected samples 
described in $\S$ \ref{hbmorph}.  For both the samples inside and outside 
the core radius, and inside and outside the elliptical boundary which splits 
the total sample approximately in half, there is no evidence (or even 
suggestion) of any radial abundance gradient.  The 3$\sigma$ upper limit
on any such gradient, within $\sim$1.3--1.5 core radii is approximately 
0.4 dex.  
This lack of any abundance gradient in And~III is consistent with the results 
for And~I and And~II (Papers I and II) and for the Galactic dSphs -- in no
dSph system has a radial abundance gradient been definitively established,
though the Galactic dSph Fornax may be an exception \citep{SHB00}.

\subsection{The Abundance Spread in And III} \label{abund_spd}

In $\S$ \ref{clr_wid} we showed that the red giant branch of And~III has
a small, but nevertheless real, intrinsic color width.  We now use the 
abundance calibration developed in the previous section to convert this
intrinsic color width into an abundance distribution\footnote{The lack of 
evidence for any intermediate-age population in And~III allows this interpretation (cf.\ Carina, \citet{TSH00} and Leo~I, \citet{CG99}).}.  
First, by 
applying the appropriate calibration to each of the 92 individual stars
in the six upper giant branch $V$ magnitude bins used for the mean 
abundance determination, we can generate 92 individual observed [Fe/H] 
values.  The distribution of these observed [Fe/H] values is shown in
Fig.\ \ref{abund_dist}.  Also shown in this figure are the observed abundance
distributions, derived in a similar fashion, for And~I and And~II from 
Paper~II\@.  The And~III distribution is approximately gaussian in shape, 
and its narrow width stands in clear contrast to the broader distributions
of And~I and And~II\@.  

\placefigure{abund_dist}

The And~III distribution can be characterized in a
number of ways: the standard deviation $\sigma_{obs}$([Fe/H]) is 0.15 dex,
the interquartile range is 0.19 and the range of the central two-thirds of
the sample is 0.32 dex.  The apparent full abundance range is $\sim$0.6, from
approximately --2.2 to --1.6 dex.  The upper limit would appear to be 
relatively secure but the lower limit might be affected by the loss of
sensitivity to metallicity of broad-band colors at low abundances, i.e.\ it
is difficult to rule out the existence of And~III stars with abundances
below this limit.  Nevertheless, as is readily  apparent from Fig.\ 
\ref{abund_dist}, these And~III values are notably smaller
than the equivalent quantities for And~I and And~II (cf.\ Paper~II)\@.  
Indeed for And~I and II, the contributions
to the observed distributions from the photometric errors are negligible,
and thus those distributions could be considered effectively intrinsic
(cf.\ Paper~II)\@.  However, this is not as obviously the case for And~III\@.

Table \ref{error_tab} shows that for the magnitude range under consideration,
the photometric errors in the F450W--F555W colors are 0.026 mag or less.
We conservatively adopt $\sigma_{err}$(F450W--F555W) = 0.03 mag.  Given the
slope of the F450W--F555W to $B-V$ transformation and the mean 
color of the red giants in the sample, this adopted photometric error 
corresponds to $\sigma_{err}$($B-V$) = 0.038 mag, or using the mean slope
of the abundance calibrations, to $\sigma_{err}$([Fe/H]) = 0.10 dex.
Subtracting this value in quadrature from the observed standard deviation
then gives $\sigma_{int}$([Fe/H]) = 0.12 for the And~III intrinsic abundance
distribution.  

This value is somewhat smaller than the limits given by ADCS93 who found
0.16 $\leq$ $\sigma_{int}$([Fe/H]) $\leq$ 0.24 from their
ground-based $V$ and $I$ photometry.  Taken at face value, the 
dispersion derived here is one of the smallest intrinsic abundance
dispersions known for a dSph galaxy (cf. \citet{MM98}).  However, we caution 
against placing too much weight on the exact value
of the And~III abundance dispersion, though it is
evident from Fig.\ \ref{abund_dist} that $\sigma_{int}$([Fe/H]) for 
And~III is not large.  Specifically, as noted above, the colors of old red 
giants are certainly metallicity dependant, but as the abundance 
is lowered below --2.3, or thereabouts, the dependance on color is much 
reduced.  Thus, it is potentially possible that a significant abundance range 
may be hidden by the color degeneracy of very metal-poor red giant stars.  Nevertheless, Fig.\ \ref{abund_dist} does not suggest this is obviously the 
case for And III\@.  

However, amongst the lower luminosity Galactic dSphs, i.e.\ those most 
comparable in luminosity to And~III, there are clear
disparities between inferences from c-m diagrams and from spectroscopy
of individual stars.  For example, for Draco (cf.\ Fig.\ \ref{draco}),
Ursa Minor, Sextans and Carina, there is only limited evidence for intrinsic
giant branch color widths in any of these systems \citep{PS79, OA85, NBS93,
TSH94}.  This is due at least in part to the relatively small numbers of 
red giants 
in these systems, and to the difficulty of deciding the membership, based on 
photometry alone, of potential members that deviate from the principal 
sequences.  For the case of Carina it appears that the age range present in
this dSph may have 
conspired with a metallicity range to produce the relatively narrow upper 
giant branch \citep{TSH00}.  However, in all cases spectroscopic observations 
indicate the presence of significant abundance ranges in these systems.  
\citet{NBS93} gives $\sigma_{int}$([Fe/H]) $\approx$ 0.30 for Draco and $\approx$0.27 for Ursa 
Minor, while \citet{NBS932} give $\sigma_{int}$([Fe/H]) = 0.19 $\pm$ 0.02 for 
Sextans and \citet{TSH00} give $\sigma_{int}$([Fe/H]) $\approx$ 0.25 for 
Carina.  

These values
all exceed our photometric value of $\sigma_{int}$([Fe/H]) for And~III and
suggest that a spectroscopic survey of member red giants is needed to
fully constrain the And~III abundance distribution.  In this context it
is worth noting that for And~II, where $\sigma_{int}$([Fe/H]) is evidently
large, the photometric abundance distribution (Paper~II) agrees well with
spectroscopic determination \citep{CO99}.  If the relatively narrow And~III
abundance distribution is confirmed via spectroscopy, then the implication 
would be that And~III retained relatively little 
of the enrichment products generated during its evolutionary history.  This 
appears also to be the case for the Galactic dSph Leo~I where \citet{CG99} 
have inferred a narrow abundance distribution from their analysis of a
WFPC2 based c-m diagram, despite the extended star formation history of this 
dwarf.  These results underline the fact that the intrinsic abundance
distribution is a vital component in any analysis of the evolutionary
history of dwarf galaxies.

\subsection{The Age(s?) of And III} \label{age_sect}

The occurrence of blue HB stars and RR Lyrae variables in the c-m diagrams
of And~III (cf.\ Fig.\ \ref{cmd1}, Fig.\ \ref{cmd2}) unambiguously indicates 
the presence of a population in this dSph whose age is essentially the same
as that of the Galactic halo globular clusters.  This result comes as
no surprise since in every dSph system where sufficient data exists, there
is evidence for the presence of such old (age $\geq$ 10 Gyr) populations.
Given the results of \citet{HE02}, this statement now also includes the
dSph Leo I, despite its dominant comparatively young population (e.g.\ 
\citet{CG99}).  For And~III, since the BHB stars are $\sim$10\%
of the total HB population (cf.\ $\S$ \ref{hbmorph}), this percentage can be
taken as a lower limit on the size of the old population in And~III\@.
On the other hand, ADCS93 have demonstrated that And~III does not 
contain any significant population with ages considerably less than 
$\sim$10 Gyr.  Based on the presence of a small number of stars brighter than 
the red giant branch tip in a field-subtracted $I$-band luminosity function, 
these authors concluded that the And~III population fraction with
ages between $\sim$3 and 10 Gyr was only 10 $\pm$ 10 per cent.
There was no indication of any yet younger stars.

Nevertheless, despite the presence of a limited number of BHB stars, the
bulk of the And~III HB population lies to the red of the instability
strip, and is thus potentially of younger age than most of the Galactic
globular clusters.  The actual age difference, however, is not easily 
quantified.  As discussed in \citet{RY01} (see also \citet{YW99, YW01}),
the latest synthetic HB models suggest an increased sensitivity of HB
morphology to age, as compared to earlier models (e.g.\ \citet{LDZ94}).
Indeed as Fig.\ 9 of \citet{RY01} suggests the age difference required
for a significant HB morphology shift is now approximately half that implied
by the earlier models.  Using this figure together with the mean And~III
abundance of $<$[Fe/H]$>$ $\approx$ --1.9 (cf.\ $\S$ \ref{abund}), the 
dominant red morphology of the And~III horizontal branch suggests that the
bulk of the population in this dSph is $\sim$3 to 3.5 Gyr younger than the
Galactic inner halo globular clusters, and $\sim$2 Gyr younger than those
globular clusters with R$_{GC}$ exceeding 8 kpc (cf.\ Fig.\ 9 of 
\citet{RY01}).  This age difference is consistent with the results of
ADCS93 who found only a small number of stars above the And~III red giant 
branch tip and thus inferred a small intermediate-age population fraction: 
only if the age difference was greater would larger numbers of such stars be 
expected.

What evidence is there to support this ``age is the second parameter''
interpretation of the And~III HB morphology?  First, there is the 
consistency between red HB morphologies and the ``young'' mean ages determined
from main sequence turnoff photometry for a number of Galactic dSphs.
As noted in the Introduction, for Carina, Leo I, Leo II and Fornax, there
is a direct observed connection, via main sequence turnoff photometry, 
between a mean age younger than that of the Galactic globular clusters,
and a red HB morphology.  This age -- horizontal 
branch morphology connection is also reinforced by Ursa Minor where the 
strong blue HB morphology in this dSph goes with an age indistinguishable 
from that of the globular clusters of the inner Galactic 
halo \citep{OA85,MB99}.  Second,
recent data for pairs of Galactic globular clusters with similar abundances 
but different HB morphologies are revealing age differences from main
sequence turnoff photometry (e.g.\ \citet{GN90, ST99, BF01, RY01} that are 
consistent with those inferred from the new HB models 
(e.g.\ \citet{RY01, CB01}).  Thus, while acknowledging that for Galactic
globular clusters additional parameters such as cluster density, rotation,
etc, may also influence HB morphology, the available evidence does support 
our interpretation of the And~III HB morphology in terms of a younger age.
Confirmation of this interpretation via photometry of the And~III main 
sequence turnoff is, unfortunately, beyond currently available resources.

In $\S$ \ref{hbmorph} we argued that the And~III HB morphology is similar
to, or even perhaps somewhat redder than, the HB morphology of Draco.
Consequently, as for And~III, we infer from Fig.\ 9 of \citet{RY01} that the
dominant population of Draco should be $\sim$3 Gyr younger than the 
majority of Galactic globular clusters.  The results of \citet{CG98}, however,
appear to contradict this assertion.  These authors, using WFPC2 
observations that reach well below the main sequence turnoff in Draco, find
that the age of the dSph is nominally 1.6 $\pm$ 2.5 Gyr {\it older}
than the Galactic halo globular clusters M68 and M92.  Modulo possible
calibration uncertainties (cf.\ $\S$ 3.2 of \citet{CG98}) and uncertainties
in the cluster fiducials against which the comparison was made (cf.\ the
discussion in $\S$ 1 of \citet{RY01}), the \citet{CG98} result is 
$\sim$1.8$\sigma$ away from our HB morphology based age for Draco.
It is difficult
to decide the significance of this contradiction.  As \citet{CG98} themselves
admit, the WFPC2 Draco observations are not of the highest quality and
their sample of turnoff stars is limited by WFPC2's small field.
A new investigation of this question based on high quality data, perhaps
obtained with the new generation 8-10m class telescopes, is clearly
warranted.  

Two other recent papers have also investigated the star formation history
of Draco.  \citet{AP01} have used data obtained from a 2.5m telescope
in their analysis.  While these data cover a wide field and thus provide
a large sample of Draco stars, the precision of the photometry in the
vicinity of the main sequence turnoff and fainter is limited.  This
restricts the degree to which they can constrain the age of the bulk
of the population in Draco.  They confirm that the bulk of the stars are
older than 10 Gyr, but cannot distinguish if ``the most conspicuous
population is 12--13.5 or 13.5--15 Gyr old'' \citep{AP01}.  The other
paper is that of \citet{AD02}, which analyses the same WFPC2 data set as
\citet{CG98}.  Despite the use of sophisticated c-m diagram modelling
techniques, \citet{AD02} concludes only that the bulk of the star 
formation in Draco occurred at early times (ages $\geq$11 Gyr); it is not
possible to establish whether this corresponds to the age of the Galactic
globular clusters, or not.  Thus we conclude that an age for Draco 
$\sim$3 Gyr younger than that of the Galactic globular clusters cannot be 
conclusively ruled out at this time.

\section{Discussion}   \label{discuss}

With the analysis of our And~III observations, we now have c-m diagrams
for all three of the original (cf.\ \citet{VB72, VB74}) dSph companions to
M31\@.  These c-m diagrams are shown in Fig.\ \ref{all3}.  All three 
systems have dominant red HB morphologies which we have interpreted as 
indicating the presence of substantial populations in each dSph 
that are younger than the Galactic globular clusters (cf.\ $\S$ \ref{age_sect}, 
Paper I, Paper II)\@.  Only And~II, however, possesses a definite 
intermediate-age population (see Paper~II and references therein).  

\placefigure{all3}

The strong red HB morphologies for the And dSphs are reminiscent of the
dominant red HB morphologies found among the Galactic dSph systems, where
the blue HB of Ursa Minor is the only counter example among the Galaxy's
nine dSph companions.  In this respect then, the two sets of dSph 
companions are similar.  They are also similar in that they obey
the same relations between mean metallicity, luminosity, central 
surface brightness
and length scale (e.g.\ \citet{NC92, NC99}).  The major {\it difference}
between the two systems is the comparative lack of intermediate-age populations
in the M31 dSphs.  Among the Galaxy's companions, Fornax, Leo~I and Carina,
for example, all have significant populations with ages less than a few
Gyr, or even younger, whereas such populations are completely unknown 
among the M31 dSph companions.  This statement can be extended to include 
the recently
discovered systems And V, VI and Cas (And VII), since ground-based
c-m diagrams for the red giant branch in these systems have not revealed
any prominent upper-AGB star populations \citep{AD98, AJ99, HS99, GG99}.  
Such upper-AGB stars 
are, however, seen in the c-m diagrams of two M81 Group dSphs, neither 
of which is obviously associated with any large galaxy \citep{NC98}.  
It then seems likely that we are seeing an environmental effect in the
comparative lack of intermediate-age stars in the M31 dSphs.  Some process 
or processes (e.g.\ ram pressure stripping in a hot gaseous corona, and/or supernovae-driven galactic winds and/or a high-UV flux from the bulge of M31) seems to have affected the M31 dSphs, removing their gas and largely terminating their star formation $\sim$10 Gyr ago.  Evidently this process or processes wasn't as strong for the Galaxy and its dSph companions.

In this context it is worth recalling that among the Galactic dSphs, there
is a tendency for the more distant systems to have larger and younger
intermediate-age populations (e.g.\citet{VB94}).  Although the size of
any intermediate-age population in the M31 dSphs is less than that of
most of the Galaxy's dSph companions, there is a suggestion here of a similar 
radial trend.
Of the three M31 systems studied so far, it is the most radially distant dSph
And~II that contains a definite intermediate-age population.  
Further, the intermediate distance system And~III has, based on its 
redder HB morphology despite a lower mean abundance, a mean age somewhat 
younger than that of the innermost system And~I\@.  It will be interesting 
to see if this possible trend is confirmed when similar analyses become 
available for the other M31 dSphs -- And~V, And~VI and Cas (And~VII).  
The latter two objects are of particular interest in this context since their 
distances from the center of M31 exceed that of And~II\@. 

One further point deserves mention.
As noted in the Introduction, \citet{BR00} have claimed an association
between an H {\small I} cloud and And~III\@.\footnote{See also
the additional information contained in the {\it Note added in proof} in
\cite{BR00}.}  The gas cloud has a heliocentric radial velocity 
of --341 $\pm$ 6 km s$^{-1}$, which is in good agreement with the optical
velocity of And III (V$_{r}$ = --351 $\pm$ 9 km s$^{-1}$, \citet{CMSO00}).
If the cloud is at approximately the same distance as And III, then it has
an H~{\small I} mass of $\sim$10$^{5}$ M$_{\sun}$ and is centered about 
3.7 kpc from the center of And III\@.  Given that the (geometric mean)
limiting radius of And III is approximately 1.35 kpc \citep{NC92}, there is
little overlap between the H {\small I} cloud and the optical extent of the
galaxy.  Clearly, as the color-magnitude diagrams of Figs.\ \ref{cmd1} and
\ref{cmd2} show,
the apparent presence of the H {\small I} cloud has had little influence
on the star formation history of And~III -- there is simply no indication
of any young stars in this dSph.

This situation is reminiscent of those for the dSphs Leo~I, Sextans
and Sculptor.  All three dSphs have claimed associations with H {\small I} 
clouds or complexes (\citet{BR00}, but see also \citet{LY00}; \citet{CB98}, 
\citet{CC99}).  The bulk of the stellar population of Leo~I has an age in the
range of 1--7 Gyr, with few, if any, stars significantly younger \citep{CG99},   while Sextans and Sculptor have no significant populations younger than a few 
Gyrs or more \citep{MM91,HK99}.  Thus, as for And~III, the apparent presence of nearby 
H {\small I} has not had any recent influence on the star formation in these
systems.  This circumstance contrasts with that for the dSph/dIrr systems 
LGS~3 and
Phoenix.  Here the H~{\small I} is coincident (LGS 3, \citet{YL97}) or
near-coincident (Phoenix, \citet{GMD01, JSG99}) with the optical galaxy, and 
both dwarfs have had star formation within the past $\sim$100 Myr or so
(e.g.\ \citet{HSG00, AG97, BM01}).  Apparently, assuming the H~{\small I}/dSph 
associations are real, it seems that possible evolutionary outcomes for such 
gas clouds include no star formation at all, star formation ceasing at a 
variety of past times, and on-going star formation.  The individual outcomes 
are presumably dependant on the density and dark matter content of the clouds, 
on the potential-well depth of any associated dwarf galaxy and on environmental 
influences.

\acknowledgements
The authors are grateful to Dr. Bart Pritzl for helpful input to the
discussion of the variable star candidates. 
 This research was supported in part
by NASA through grant number GO-07500 from the Space Telescope Science
Institute, which is operated by AURA, Inc., under NASA contract NAS 5-26555.

\clearpage
\figcaption[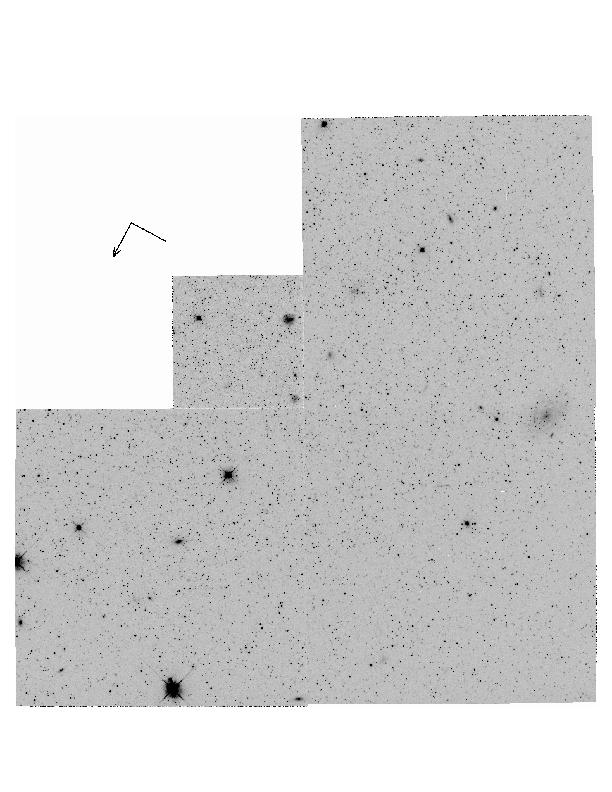]{A mosaic of the And III WFPC2 field made from 
the combination of four 1200~s F555W exposures.  North is indicated by the 
direction of the arrow and East by the line.  Both indicators are 10$\arcsec$ 
in length.    \label{wfpc2pic} }

\figcaption[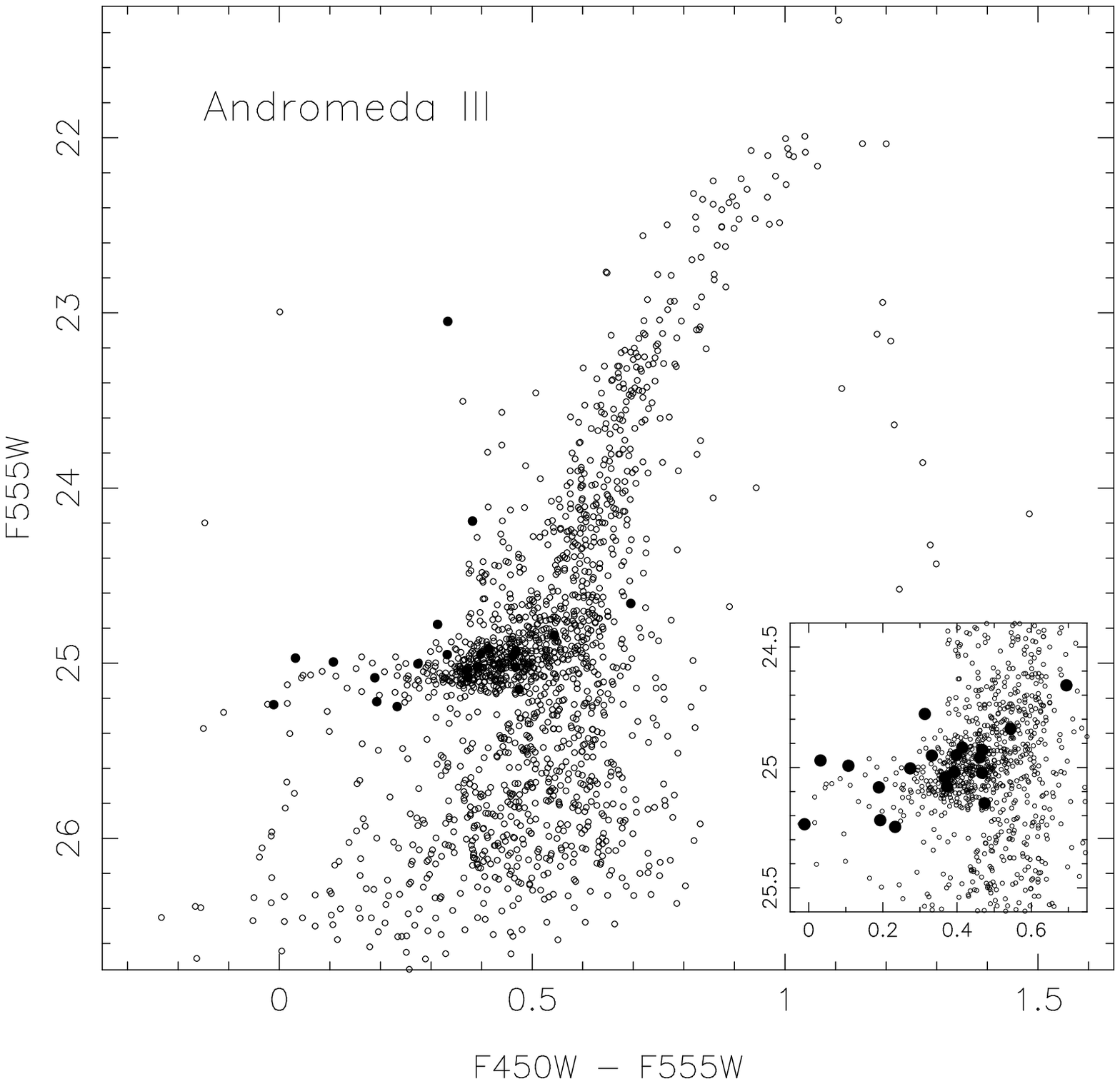]{A color-magnitude diagram for Andromeda III\@.
The photometry is on the F450W, F555W system of \citet{H95B}\@.  Charge transfer
efficiency corrections from \citet{WHC99} have been applied.  The filled
symbols are stars which show large magnitude or color differences between
the two sets of observations.  Most are likely to be And~III variable stars.
The insert box enlarges a region around the horizontal branch.
\label{cmd1} }

\figcaption[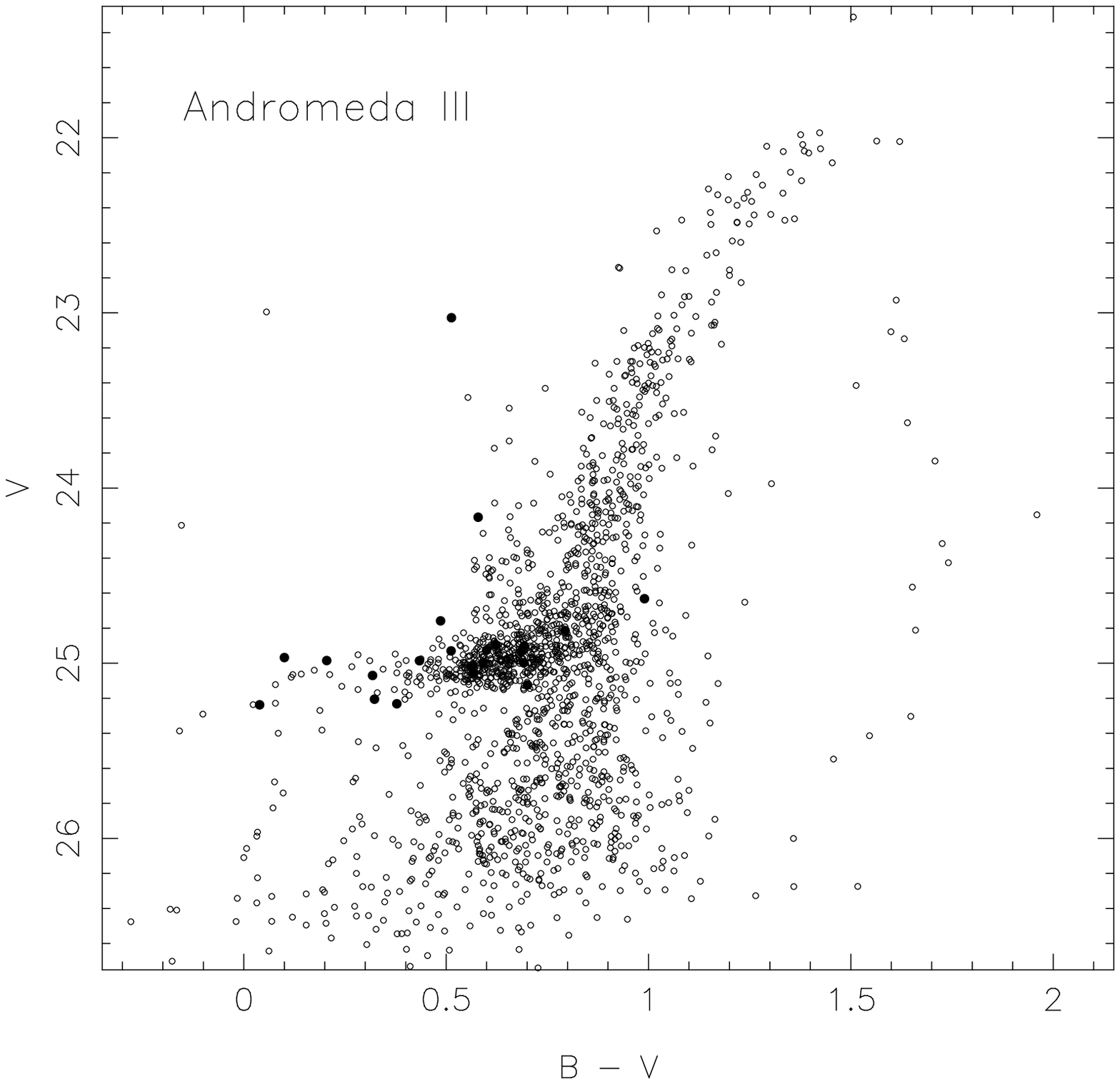]{A color-magnitude diagram for Andromeda~III
on the standard $B$, $V$ system.  The candidate variable stars are 
plotted as filled symbols. \label{cmd2} }

\figcaption[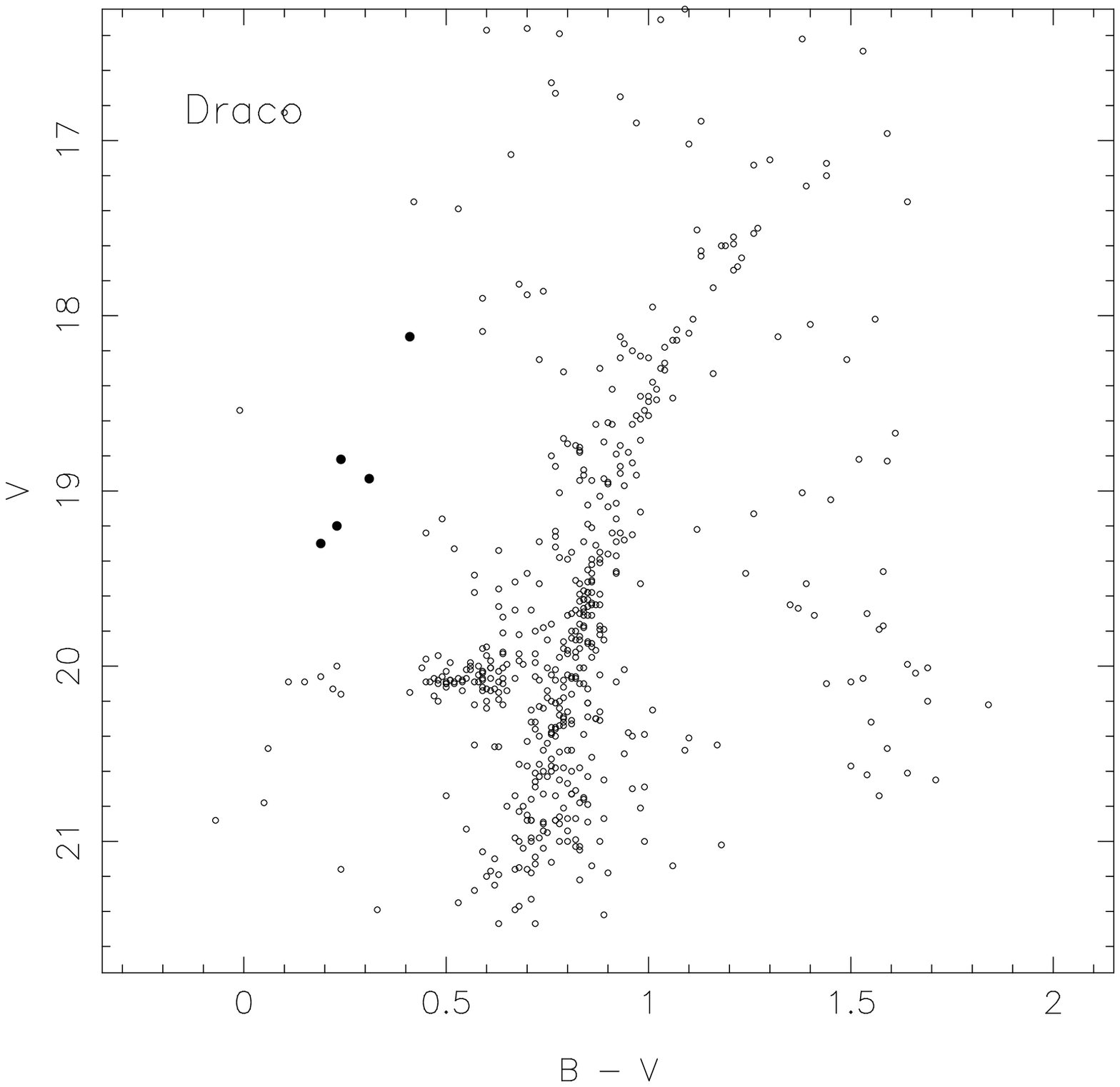]{A color-magnitude diagram for the Galactic
dSph Draco using the photometry of \citet{PS79}.  The filled symbols are
Draco Anomalous Cepheid variables plotted at values of $<V>$ and $<B>-<V>$
tabulated by \citet{BP01}.  \label{draco} }

\figcaption[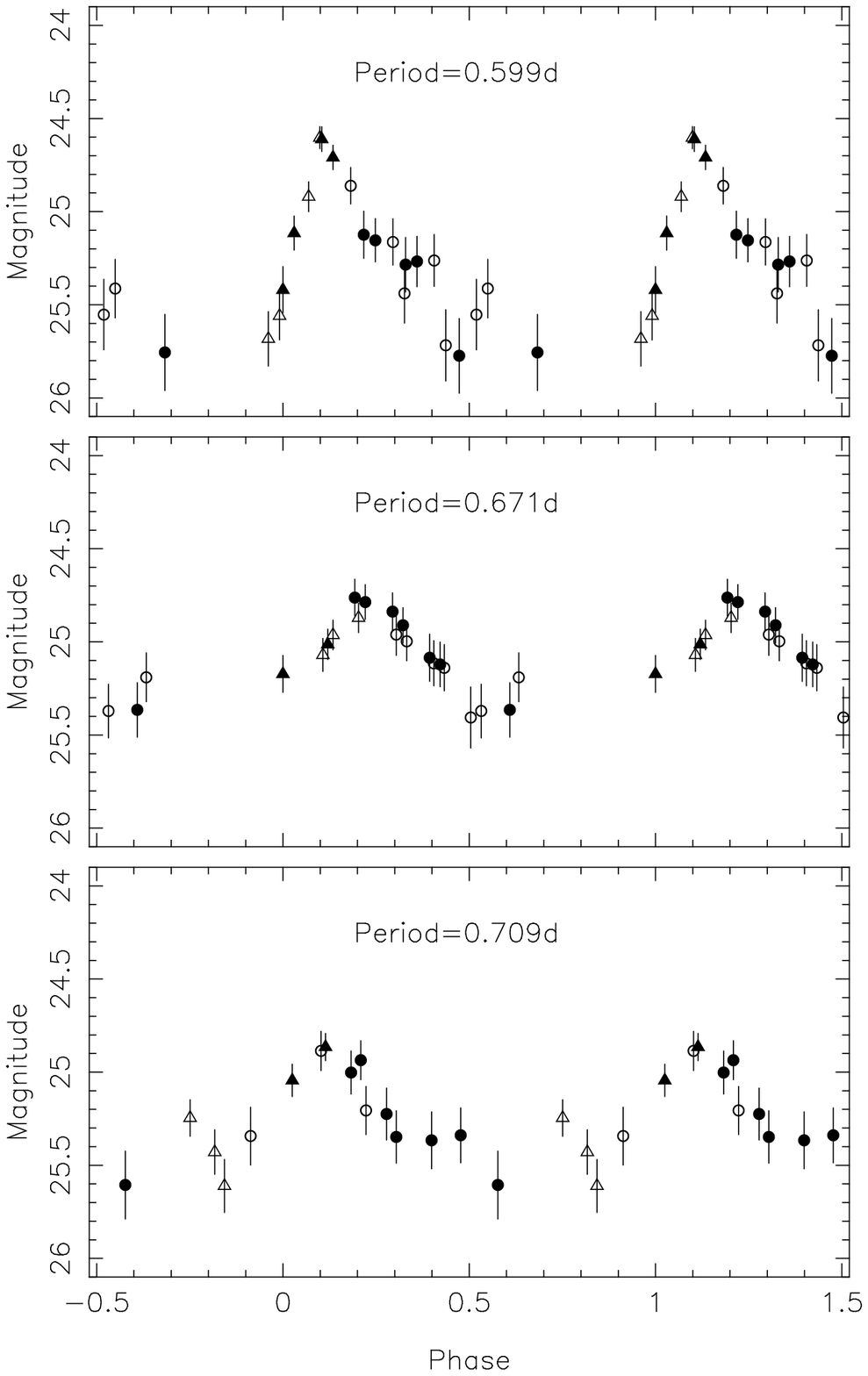]{Light curves for 3 And III RR Lyrae variables.
These stars come from the candidate variables identified in Figs.\ \ref{cmd1}
and \ref{cmd2}.  Two cycles are plotted for each star, with the filled and
unfilled symbols representing the two sets of observations.  F450W magnitudes
are shown as circles while F555W magnitudes are shown as triangles.  A 
constant F450W--F555W color of 0.20 mag has been assumed to place the F555W observations with those for F450W\@.  The error bars are those from photon
statistics and the period adopted for each star is given in the panels.
\label{rr_fig} }

\figcaption[DaCosta.fig6.eps]{Light curves for 1 definite (WF2-1398; upper 
panel) and 1 probable (WF2-1710; lower panel)
Anomalous Cepheid variables in And~III\@.  Two cycles are plotted for each 
star, with the filled and unfilled symbols representing the two sets of 
observations.  F450W magnitudes
are shown as circles while F555W magnitudes are shown as triangles.  For 
WF2-1398, a constant F450W--F555W color of 0.20 mag has been assumed to place the F555W observations with those for F450W\@.  For WF2-1710 the color values
used were 0.15 and 0.40 mag, respectively.  The error bars are those from 
photon statistics and the period adopted for each star is given in the panels.
\label{ac_lght} }

\figcaption[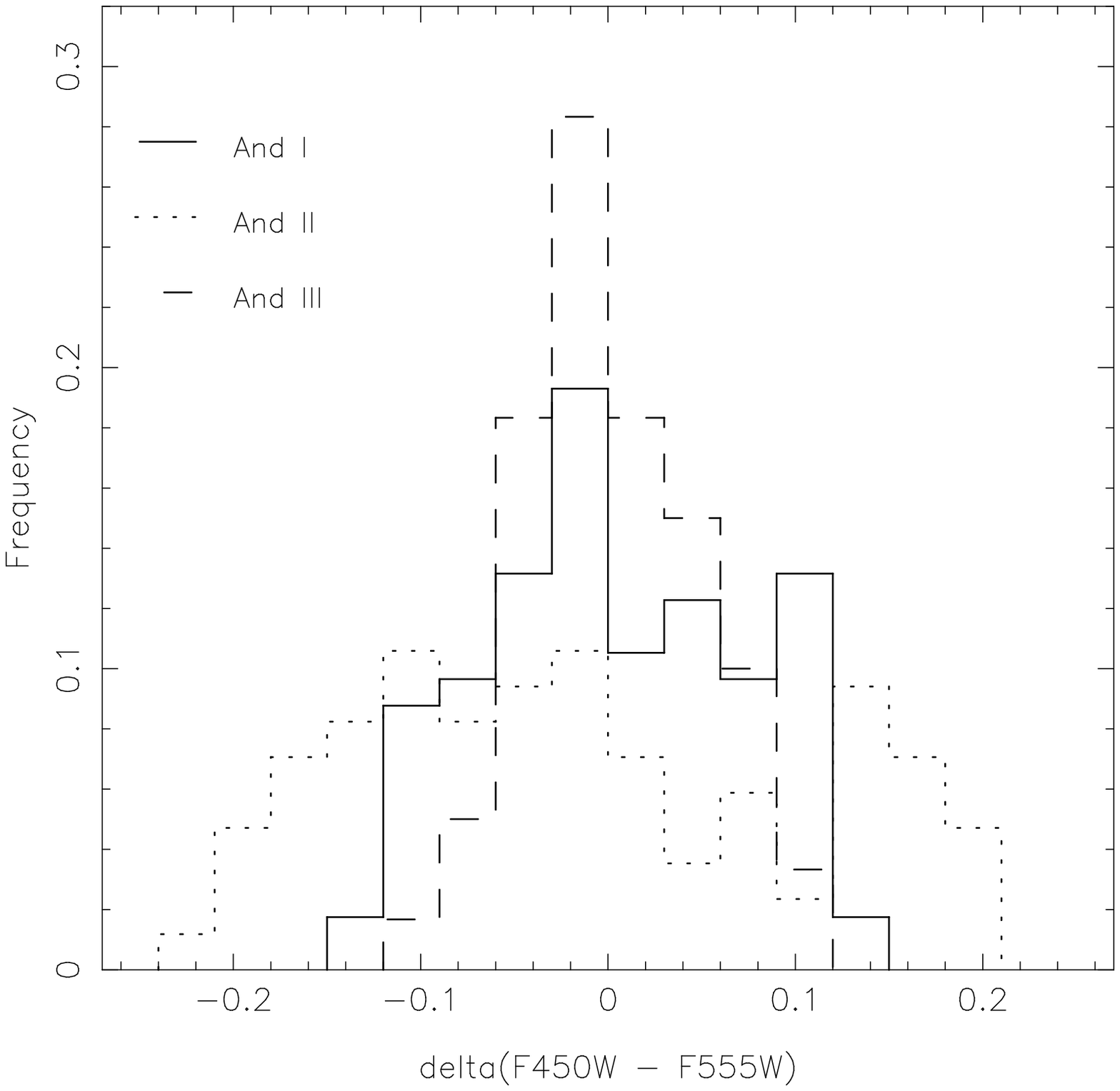]{Normalized histograms of observed 
F450W--F555W residuals from
the mean giant branches for And~I (solid line), And~II (dotted line)
and And~III (dashed line).  The sample sizes are 114, 85 and 60 stars,
respectively.  For all three dSphs, the photometric errors for the 
F450W--F555W colors are 0.03 mag, the bin size, or less. \label{delta_col} }

\figcaption[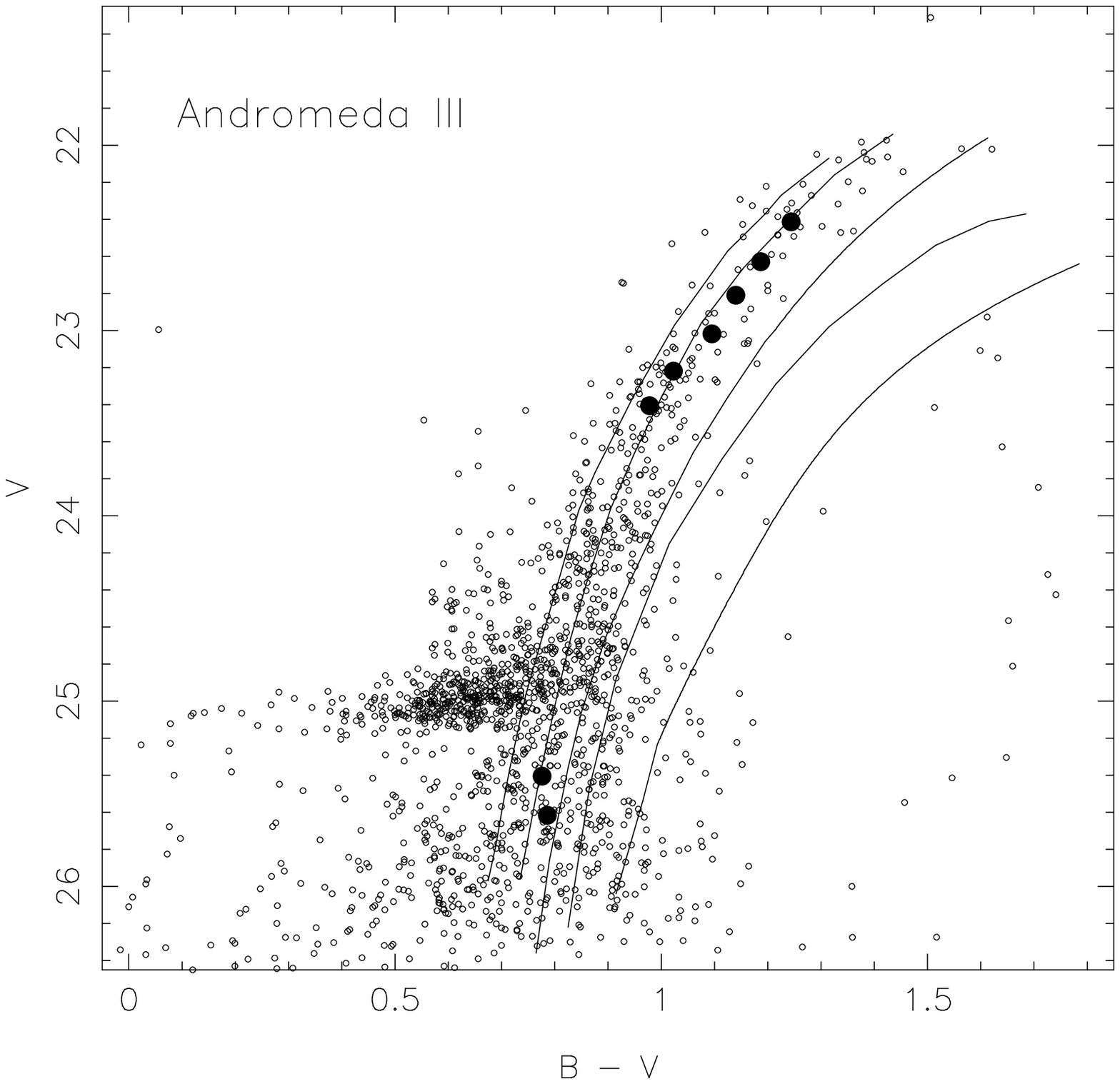]{The color-magnitude diagram from 
Fig.\ \ref{cmd2}, excluding the candidate variable stars, is shown superposed 
with red giant branches from the standard globular clusters M68 ([Fe/H] = 
--2.09), M55 ([Fe/H] = --1.82), NGC~6752 ([Fe/H] = --1.54), NGC~362 ([Fe/H] = 
--1.28), and 47~Tuc ([Fe/H] = --0.71).  The globular cluster data have been
placed in this diagram using $(m-M)_{V}$ = 24.56 and $E(B-V)$ = 
0.055 for And~III\@.  The filled symbols give the mean giant branch colors
for bins of $\pm$0.1 mag in $V$.  \label{gbcmd} }

\figcaption[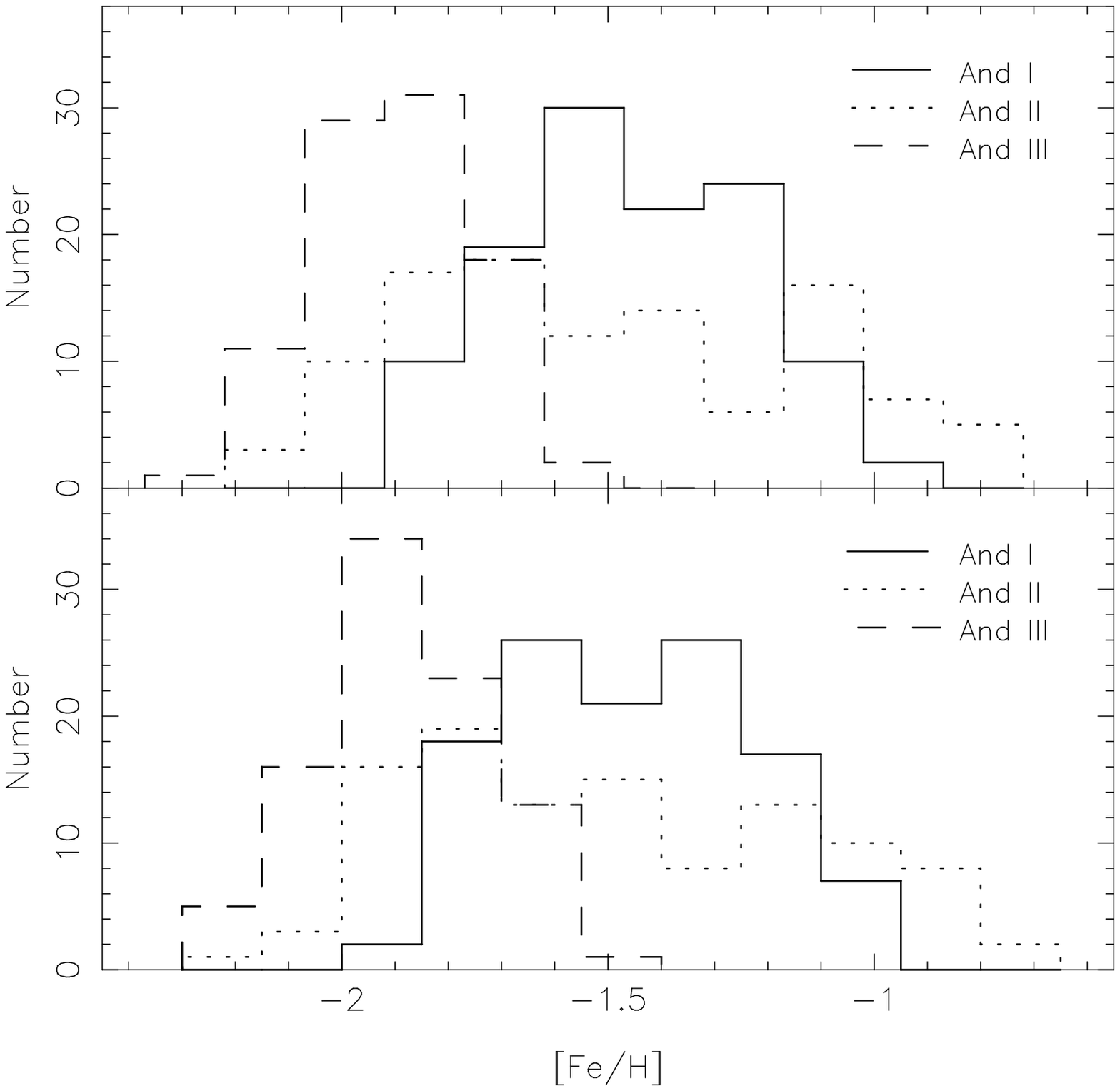]{Histograms of observed abundance distributions
derived from red giant branch colors for And~I (solid line), And~II (dotted
line), both from Paper~II, and And~III (dashed line).  The sample sizes are
117, 108 and 92 stars, respectively, and the average error in the individual
[Fe/H] values is 0.10 dex, or less.  The bin size is 0.15 dex and the two
panels show the effect of moving the centers of the histogram bins by 0.07
dex. \label{abund_dist} }

\figcaption[DaCosta.fig10.eps]{Color-magnitude diagrams for And~III (upper),
And~II (middle) and And~I (lower) from our WFPC2 programs.  \label{all3} }

\clearpage
\begin{deluxetable}{lcccc}
\tablecolumns{5}
\tablewidth{0pt}
\tablenum{1}
\tablecaption{Photometric Errors.}
\tablehead{\colhead{F555W}&\colhead{Mean Error}&\colhead{Mean Error in}
&\colhead{F450W}&\colhead{Mean Error}\\
&\colhead{in F555W}&\colhead{F450W--F555W}&&\colhead{in F450W}}
\startdata
21.5--23.0 & 0.013 & 0.018 & 22.0--24.4 & 0.022\\
23.0--24.0 & 0.016 & 0.026 & 24.4--24.9 & 0.030\\
24.0--24.5 & 0.024 & 0.040 & 24.9--25.4 & 0.044\\
24.5--25.0 & 0.035 & 0.058 & 25.4--25.8 & 0.054\\
25.0--25.4 & 0.043 & 0.068 & 25.8--26.2 & 0.078\\
25.4--25.8 & 0.060 & 0.099 & 26.2--26.6 & 0.111\\
25.8--26.1 & 0.078 & 0.136 & 26.6--26.9 & 0.145\\
26.1--26.4 & 0.107 & 0.180 & 26.9--27.2 & 0.187\\

\enddata 
\label{error_tab}
\end{deluxetable}

\end{document}